\documentclass[prc,aps,nofootinbib,twocolumn]{revtex4-1}
\usepackage{mathrsfs}
\usepackage{amssymb}
\usepackage{amsmath}
\usepackage{graphicx}

\usepackage[caption = false]{subfig}
\usepackage[normalem]{ulem}
\usepackage{xcolor}
\usepackage{hyperref}
\usepackage{slashed}
\usepackage{tikz}
\usepackage{cleveref}
\usepackage{xspace}
\usepackage{braket}

\newcommand{\R}{\mathbb{R}}

\newcommand{\dd}{\mathrm{d}}

\newcommand{\ii}{\mathrm{i}}
\newcommand{\ee}{\mathrm{e}}

\newcommand{\UNIT}[1]{\ensuremath{\,{\rm #1}}\xspace}

\newcommand{\MeV}{\UNIT{MeV}}

\newcommand{\fm}{\UNIT{fm}}

\newcommand{\erw}[1]{\left \langle #1 \right \rangle}
\newcommand{\REM}[1]{}

\definecolor{magenta}{cmyk}{0,1,0,0}

\newcommand{\matrixe}[3]{\ensuremath{ \left \langle{#1} \left| \vphantom
        {#1 #3} {#2} 
\right| {#3} \right \rangle }}
\makeindex

\begin{document}

\title{Bound State Formation in Time Dependent Potentials}

\author{Jan Rais}
\email{rais@th.physik.uni-frankfurt.de}
\affiliation{Institut f\"ur Theoretische Physik,
  Johann Wolfgang Goethe-Universit\"at,
  Max-von-Laue-Str.\ 1, D-60438 Frankfurt am Main, Germany}

\author{Hendrik van Hees}
\affiliation{Institut f\"ur Theoretische Physik,
  Johann Wolfgang Goethe-Universit\"at,
  Max-von-Laue-Str.\ 1, D-60438 Frankfurt am Main, Germany}

\author{Carsten Greiner}
\affiliation{Institut f\"ur Theoretische Physik,
  Johann Wolfgang Goethe-Universit\"at,
  Max-von-Laue-Str.\ 1, D-60438 Frankfurt am Main, Germany}

\date{\today}


\begin{abstract}
  We study the temporal formation of quantum mechanical bound states within a
  one-dimensional attractive square-well potential, by first solving the
  time-independent Schr{\"o}dinger equation and then study a time
  dependent system with an external time-dependent potential. For this
  we introduce Gaussian potentials with different spatial and temporal
  extensions, and  generalize this description also for subsequent pulses and for random, noisy potentials.  
  Our main goal is to study the time scales,
  in which the bound state is populated and depopulated. Particularly we
  clarify a likely connection between the uncertainty relation for
  energy and time and the transition time between different energy eigenstates. We
  demonstrate, that the formation of states is not delayed due to the
  uncertainty relation but follows the pulse shape of the
  perturbation. In addition we investigate the (non-)applicability of
  first-order perturbation theory on the considered quantum system. 
\end{abstract}

\maketitle


\section{Introduction}
\label{sec:Intro}

In relativistic heavy-ion collisions, as performed with the most
energetic accelerators like the RHIC or the LHC, the abundant production
of hadronic particles has been established over the years in order to
study the very facets and properties of the created hot and dense
matter. Recently, the yields of light nuclei, such as deuterons,
tritons, hyper-tritons, helium-3 or helium-4, and also of the
anti-nuclei as similar counterparts have been measured by the ALICE
collaboration at LHC \cite{ALICE:2015oer, ALICE:2015wav,Braun_Munzinger_2019}.

As rather fragile quantum bound states like, e.g., the deuteron with a
binding energy of only 2.3 MeV the immediate question comes up when and
how such states do form and appear: A rather surprising observation is
the fact that the overall experimental yields of the light nuclei and
anti-nuclei are in an obvious agreement with the calculated yields
obtained with the statistical hadronization model, which is
characterized by a chemical freeze-out temperature of
$T_{\text{ch}} = 155$ MeV and nearly vanishing net baryon density
\cite{Andronic:2010qu,Andronic:2017pug}. At such conditions of the
fireball the system is still very energy-dense and hot and thus any
potential light nuclei shall not exist as expected.

A phenomenological description for the microscopic production of
deuterons or even larger light nuclei are the coalescence models and
leads to a good agreement with data on low-temperature cluster formation
\cite{Toneev:1983cb, Nagle_sorge, Monreal_Greiner1,Neubert:2003qm,Braun_Munzinger_2019},
but also at much higher energies \cite{Botvina:2017yqz, Glassel:2021rod}. Here it is typically
assumed that the nucleons had their last interactions in the system and
if two (or more) are close in space and also close in momentum-space,
those nucleons may coalesce to a bound nuclear state \cite{Scheibl}.

In contrast to such a phenomenology, it has been shown that, e.g.,
deuterons have to be produced by three-body reactions of three nucleons
to a deuteron and a nucleon in the evolving system, and also its
potential dissociation are given by the reverse interactions fulfilling
the principle of detailed balance \cite{DANIELEWICZ1991712}. In this
respect, very recently it has been shown that within such a kinetic
transport approach, where the continuous production and subsequent
dissociation of deuterons \cite{Oliinychenko:2018ugs, Vovchenko_2020,
  Oliinychenko:2020znl, NEIDIG2022136891, Sun:2021dlz,
  Staudenmaier:2021lrg} as well as for the more massive light nuclei
\cite{NEIDIG2022136891} are incorporated, the experimental findings of
the LHC results can reasonably be described. The special role of the
conservation of the baryon and anti-baryon number also bridges the
result of the statistical hadronization description and the kinetic
description of such dissociation and regeneration of light nuclei
\cite{Vovchenko_2020, NEIDIG2022136891}.  A possible criticism is that
the formation time of such bound states like, e.g., the deuteron underlies
the Heisenberg's uncertainty relation in energy and time so that it may
scale as the inverse binding energy $ \tau \sim 1/E_{\text{B} }$ and
thus much longer than the system time.
	
In this present work a simple nonrelativistic one-dimensional quantum
system with one profound and distinct bound state and having a continous
quantum spectrum of unbounded states will be investigated being exposed
to time-dependent and spatially localized pulses. A single particle can
stay initially in the bound state or in an excited, freely moving
state. Due to the action of the time-dependent pulse, the temporal
dissociation or, alternatively, the temporal population of the bound
state can be simulated and analyzed. With this at hand one can study the
time scales, in which the bound state is being created or destroyed.
	
The paper is organized as follows: In \cref{Stat_Wave} the
one-dimensional quantum system is introduced, solving then for the
Schr\"odinger equation and to obtain the energy spectrum. The distinct
bound state is constructed similarly to a deuteron with a box-type
potential. In \cref{Time-dep} the time-dependent Schr\"odinger equation is
set up with an additional external time-dependent and localized
potential. The formal solution will be expanded in the basis of the
undisturbed quantum system. \Cref{sec:POTENTIAL} shows various results
for the reaction of the quantum particle being exposed to one single pulse
or a few pulses. With this, we will also briefly discuss the reaction of
the system to a random noise in the next \cref{sec:random}.  In
addition, the potential (non-)applicability of first-order perturbation
theory on the considered quantum system will be considered in the
following \cref{sec:pert}. Finally, in \cref{heisenberg} we intend to
clarify a likely connection between the uncertainty relation for energy
and time and the transition time between different energy
eigenstates. The formation of states is not delayed due to the
uncertainty relation, but basically follows the pulse shape of the
acting perturbation. An additional and more formal discussion is given
in \cref{Mandelstamm-Tamm} for the interpretation of the energy-time
uncertainty relation.  We close the findings of this study with a
summary and an outlook.


\section{Stationary wave function}
\label{Stat_Wave}
To address the question of the dynamical formation and destruction of a bound
state due to the influence of a time-dependent potential,
mimicking the scatterings or kicks with particles in a bath we consider the
one-dimensional motion of a single particle in the potential,
 \begin{equation}
\label{eq1}
V_0(x) = \begin{cases} \infty & \mbox{for } -\infty < x < -L, \\ 0 & \mbox{for } -L \leq x \leq -a \text{ (area 1)},  \\ -V_0 & \mbox{for } -a \leq x \leq a  \text{ (area 2)},  \\ 0 & \mbox{for } a \leq x \leq L  \text{ (area 3)},  \\ \infty & \mbox{for } L \leq x \leq \infty. \end{cases} 
\end{equation}	
In the following the energy eigenfunctions are defined as 
solutions of the time-independent Schr\"odinger equation,
\begin{equation}\label{eqstat}
\hat{H}_0 \psi_n(x)  = E_n \psi_n(x)
\end{equation}
with
\begin{equation}
\label{H_0}
\hat{H}_0 = -\frac{\hbar^2}{2m} \nabla^2 + \hat{V}_0(x).
\end{equation}
Since the Hamiltonian is symmetric under spatial reflections
$x \rightarrow -x$ we can choose the energy eigenfunctions as parity
eigenstates, fulfilling
\begin{equation*}
\psi_1^s (-x) = \psi_3^s (x) \qquad\text{and}\qquad  \psi_2^s (x) = \psi_2^s (-x)
\end{equation*}
for the symmetric and 
\begin{equation*}
\psi_1^a (-x) = -\psi_3^a (x) \qquad\text{and}\qquad  \psi_2^a (-x) = -\psi_2^a (x)
\end{equation*}
for the antisymmetric solutions. This simplifies the calculation
for constructing the wave function for
area 1 and 2 only, where area 1 is in between $-L$ and $-a$ and area 2
in between $-a$ and $a$, and implies, that the wave functions are
real. This leads to the ansatz
\begin{equation}\label{eq2}
\psi_{1,2} (x)=A_{1,2} \exp(k_{1,2} x) + B_{1,2} \exp(-k_{1,2} x),
\end{equation}
which has to fulfill the boundary and continuity conditions
\begin{align*}
\psi_{3} (L)&=\psi_{1} (-L) = 0, \nonumber\\
\psi_{1} (-a)&=\psi_{2} (-a), \nonumber\\
\psi_{2} (a)&=\psi_{3} (a), \nonumber\\
\partial_x \psi_{1}(-a)&=\partial_x \psi_{2}(-a) \nonumber\\
\partial_x \psi_{2}(a) &=\partial_x \psi_{3}(a)
\end{align*}
with 
\begin{align}\label{energy}
k_{1,n}^2 = -\frac{2m}{\hbar^2} E_n \qquad\text{and}\qquad k_{2,n}^2 = -\frac{2m}{\hbar^2} (E_n - V_0)
\end{align} 
as a result of the stationary Schr\"odinger equation, and therefore the relation
$E_n = -\frac{\hbar^2k_{1,n}^2 }{2m}$, which is to calculate the energy eigenvalues.

\begin{figure}
	\centering
	\includegraphics[width=\columnwidth]{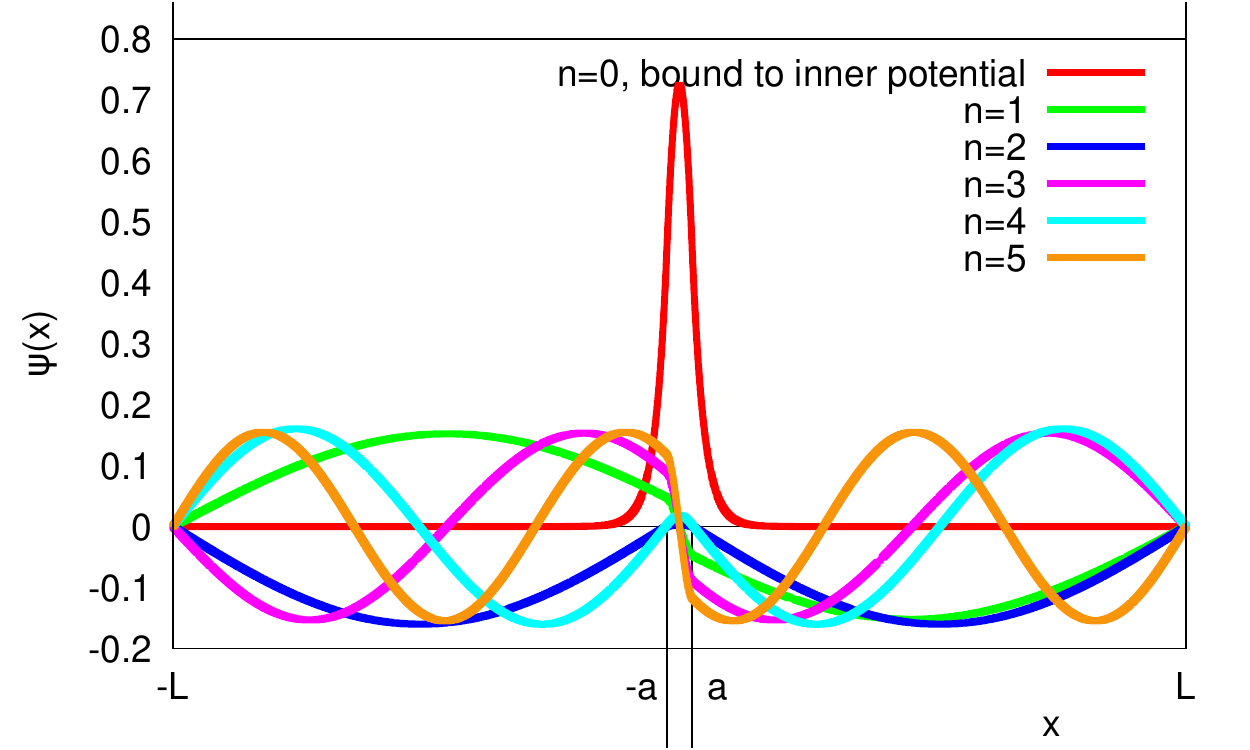}
	\caption{The first six wave functions for the double potential well. Here $2a = 1.2$ fm and $L = 100a$.}
	\label{fig:wave}
\end{figure}
\begin{figure}
	\centering
	\includegraphics[width=\columnwidth]{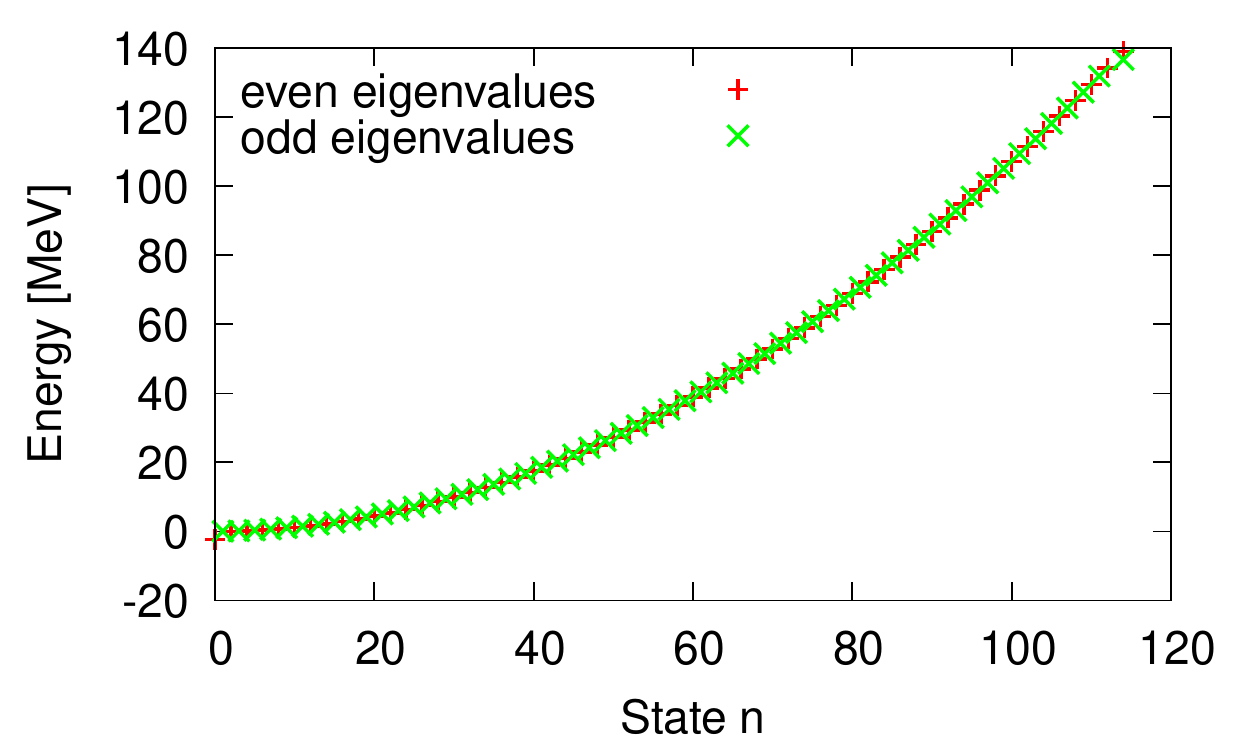}
	\caption{Energy eigenvalues for deuteron with binding energy of $-2.3$ MeV.}
	\label{fig:energy}
\end{figure}
From the boundary conditions the following solutions for symmetric and antisymmetric wave emerge,
 
 \begin{equation}\label{eq11}
 \begin{aligned}
 \psi_{1,<}^s(x) &= A \left[ \text{e}^{k_1^s x} -\text{e}^{-k_1^s (2L+x)}\right] \\
 \psi_{2,<}^s(x) &= A  \frac{ \left[ \text{e}^{-k_1^s a} -\text{e}^{k_1^s (a-2L)}\right] }{\cos(k_2^s a)} \cos(k_2^s x) \\
 \psi_{3,<}^s(x)& = A \left[ \text{e}^{-k_1^s x} -\text{e}^{k_1^s (x-2L)}\right]\\
 \psi_{1,<}^a(x) &= A \left[ \text{e}^{k_1^a x} -\text{e}^{-k_1^a (2L+x)}\right] \\
 \psi_{2,<}^a(x) &= A  \frac{ \left[ \text{e}^{-k_1^a a} -\text{e}^{k_1^a (a-2L)}\right] }{\sin(k_2^a a)} \sin(k_2^a x)\\
 \psi_{3,<}^a(x)& = -A \left[ \text{e}^{-k_1^a x} -\text{e}^{k_1^a (x-2L)}\right]
 \end{aligned}
 \end{equation}

for $E < 0$ and
\begin{equation}\label{eq12}
\begin{aligned}
\psi_{1,>}^s(x)&=A'\frac{\sin(k_{1,n}(x+L))}{\sin(k_{1,n}L)}\\
\psi_{2,>}^s(x)&=A' \frac{\sin(k_{1,n}(L-a))}{\sin(k_{1,n}L)\cos(k_{2,n}a)}\cos(k_{2,n}x)\\
\psi_{3,>}^s(x)&= A' \frac{\sin(k_{1,n}(L-x))}{\sin(k_{1,n}L)}\\
\psi_{1,>}^a(x) &=A'\frac{\sin(k_{1,n}(x+L))}{\sin(k_{1,n}L)}\\
\psi_{2,>}^a(x) &=A' \frac{\sin(k_{1,n}(L-a))}{\sin(k_{1,n}L)\sin(k_{2,n}a)}\sin(k_{2,n}x)\\
\psi_{3,>}^a(x) &= A' \frac{\sin(k_{1,n}(L-x))}{\sin(k_{1,n}L)}
\end{aligned}
\end{equation} 
for $E>0$, where $A$ and $A'$ are determined numerically to satisfy
normalization, cf. \cref{fig:wave}, and $s$ and $a$ denote parity even
and odd real valued solutions, respectively.  Finally, from the
continuity condition
 \begin{align*}
 \left.\frac{\text{d} \psi_1(x)}{\text{d} x} \right|_{-a} =  \left.\frac{\text{d} \psi_2(x)}{\text{d} x} \right|_{-a} 
 \end{align*}
the equations for the energy eigenvalues 
\begin{align}\label{eq13}
\tan(k_{2,n}^s a) = \frac{k_{1,n}^s}{k_{2,n}^s} \frac{1 +\text{e}^{2k_{1,n}^s (a-L)}}{1 -\text{e}^{2k_{1,n}^s (a-L)}}
\end{align}
for the symmetric and
\begin{align}\label{eq14}
\cot(k_{2,n}^a a) =- \frac{k_{1,n}^a}{k_{2,n}^a} \frac{1 +\text{e}^{2k_{1,n}^a (a-L)}}{1 -\text{e}^{2k_{1,n}^a (a-L)}}
\end{align}
for the antisymmetric wave functions follow, which are solved
numerically. In \cref{fig:energy} the energy eigenvalues for the first
110 eigenstates are shown, using ``deuteron parameters'', $2a=1.2$ fm,
$E_{\text{bind}} = -2.3$ MeV \cite{povh14}, resulting from choosing the
potential $V_0 =-18$ MeV and $L=100$ fm. In the calculation, $m$ is
chosen to be the reduced mass $m_{\text{p}}/2$ \cite{povh14}. Comparing
\cref{eq13} and \cref{eq14} with \cref{energy}, one realizes from
\cref{fig:energy}, that $E_n \sim n^2$ for large $n$. For small $n$ the
growth is not exactly quadratic, due to the exponential parts of
\cref{eq13} and \cref{eq14}. For the chosen parameters there is only one
``bound state'' with $E<0$.

For the following numerical calculations the energy eigenbasis up to
the $110^{\text{th}}$ state will be truncated, which corresponds to an energy cut-off of about
$140$ MeV. As will be discussed in the next section, the eigenbasis will 
serve as a restricted Hilbert space.

\section{Time dependent wave function}\label{Time-dep}

In the following we employ the energy eigenfunctions of $\hat{H}_0$ to
solve the time-dependent problem,
 \begin{equation}
\begin{split}
\label{eq10}
i\hbar \partial_t \psi(x,t) &= \hat{H} \psi(x,t) \\
&=
   \left[-\frac{\hbar^2}{2m}\partial_x^2+V_0(x) + V(x,t)\right]
   \psi(x,t),
\end{split}
\end{equation}
where $V(x,t)$ represents a time-dependent ``external potential''. Expanding
the state in terms of the $\hat{H}_0-$eigenstates $\ket{\psi_n}$,

 \begin{align}\label{eq21}
 \ket{\psi} = \sum_n c_n(t) \ket{\psi_n},
 \end{align}
it follows for the time-dependent Schr\"odinger equation 
 \begin{equation}
\begin{split}
  \ii \hbar \frac{\dd}{\dd t} \ket{\psi} &=\ii \hbar \sum_n \dot{c}_n(t)
  \ket{\psi_n} \\
  &= \hat{H}\ket{\psi} = \sum_n c_n(t) \left[E_n + \hat{V}\right]
  \ket{\psi_n},
\end{split}
 \end{equation}
where $E_n$ are the eigenvalues of $\hat{H}_0$. Multiplying with
$\bra{\psi_j}$ leads to
\begin{equation}
\ii \hbar \dot{c}_j(t) = E_j c_j(t) + \sum_{n} V_{jn}(t) c_n(t)
\end{equation}
with the matrix elements
\begin{equation}
V_{jn}(t)=\matrixe{\psi_j}{\hat{V}}{\psi_n}.
\end{equation}

\begin{figure}
	\centering
	\includegraphics[width=\columnwidth]{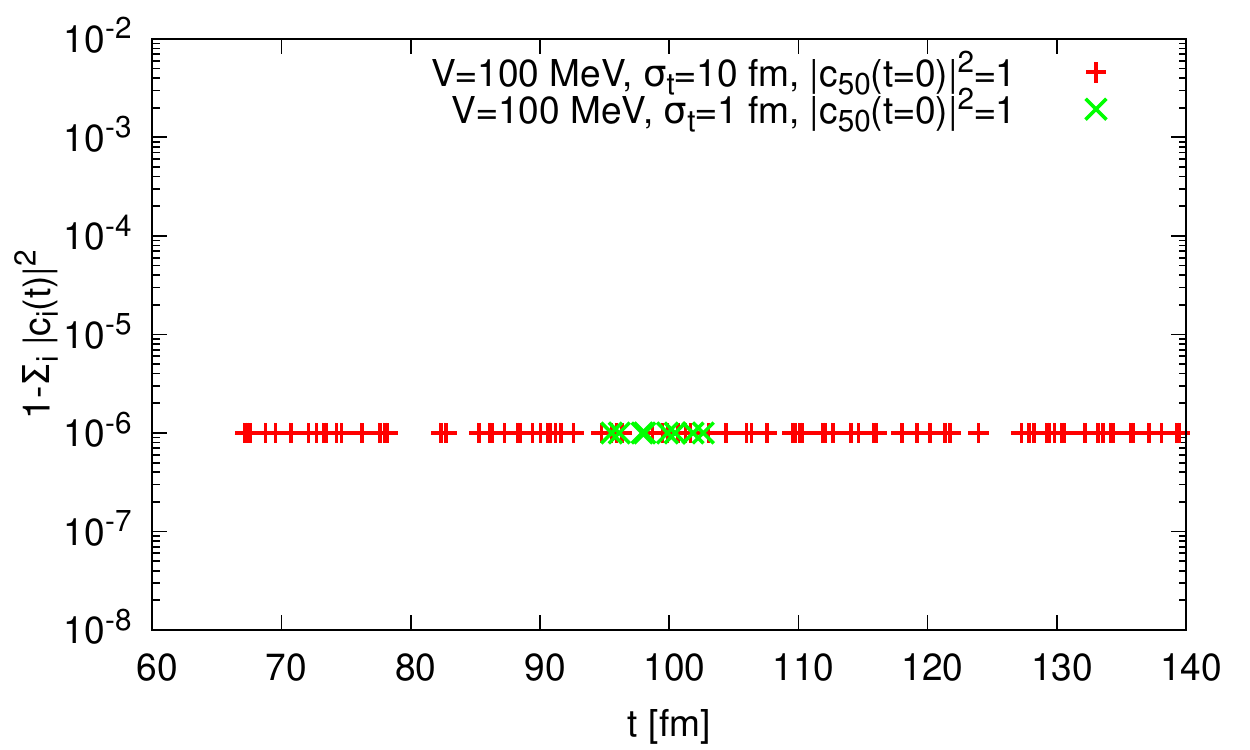}
	\caption{$1-\sum_n \vert c_n(t)\vert^2$ for a single pulse at $t=100$ fm with $\sigma_x = 1.2$ fm.}
	\label{fig:norm_t}
\end{figure}
  \begin{figure}[h]
 	\centering
 	\includegraphics[width=\columnwidth]{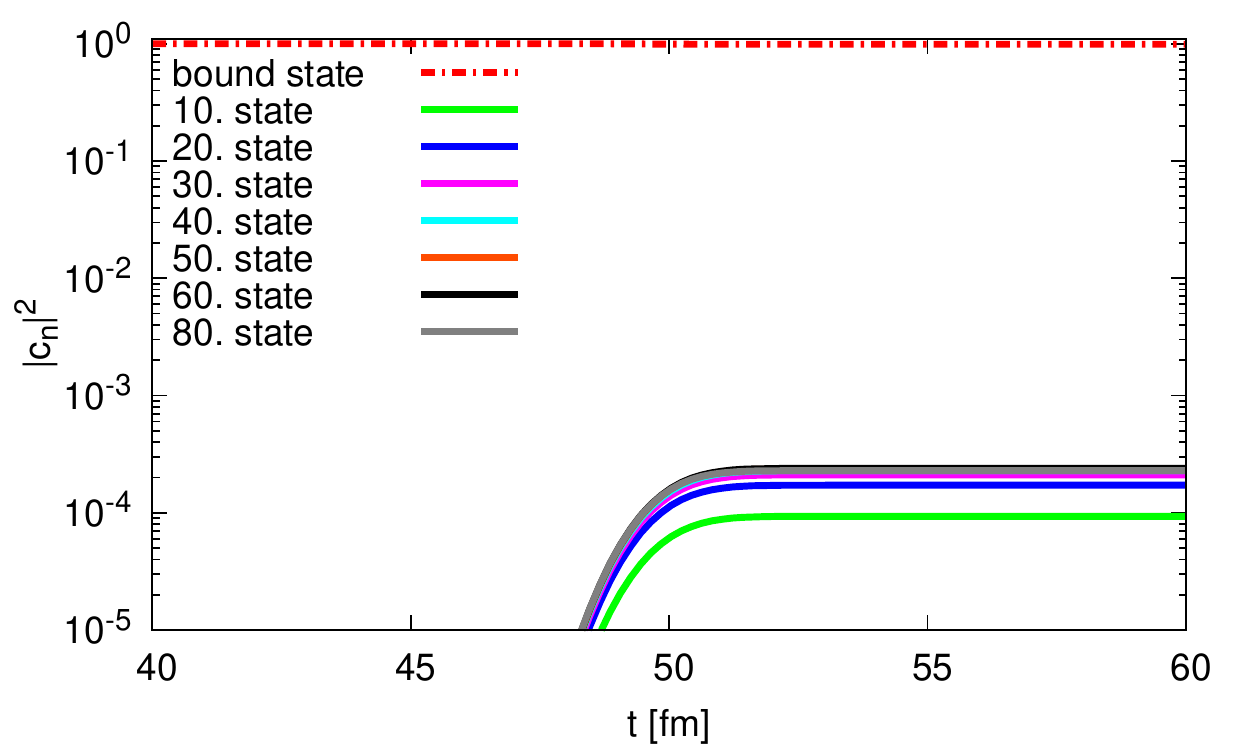}
 	\caption{$|c_n (t)|^2$ with a time dependent potential. The bound state is originally populated, $|c_0(t=0)|=1$ and $|c_{n\neq 0}(t=0)|=0$, $\sigma_t = 1 $ fm and $\sigma_x = 0.2 a = 0.12$ fm, furthermore $t_0 = 50$ fm and $V=100$ MeV.}
 	\label{fig:0state}
 \end{figure}
 \begin{figure}[h]
 	\centering
 	\includegraphics[width=\columnwidth]{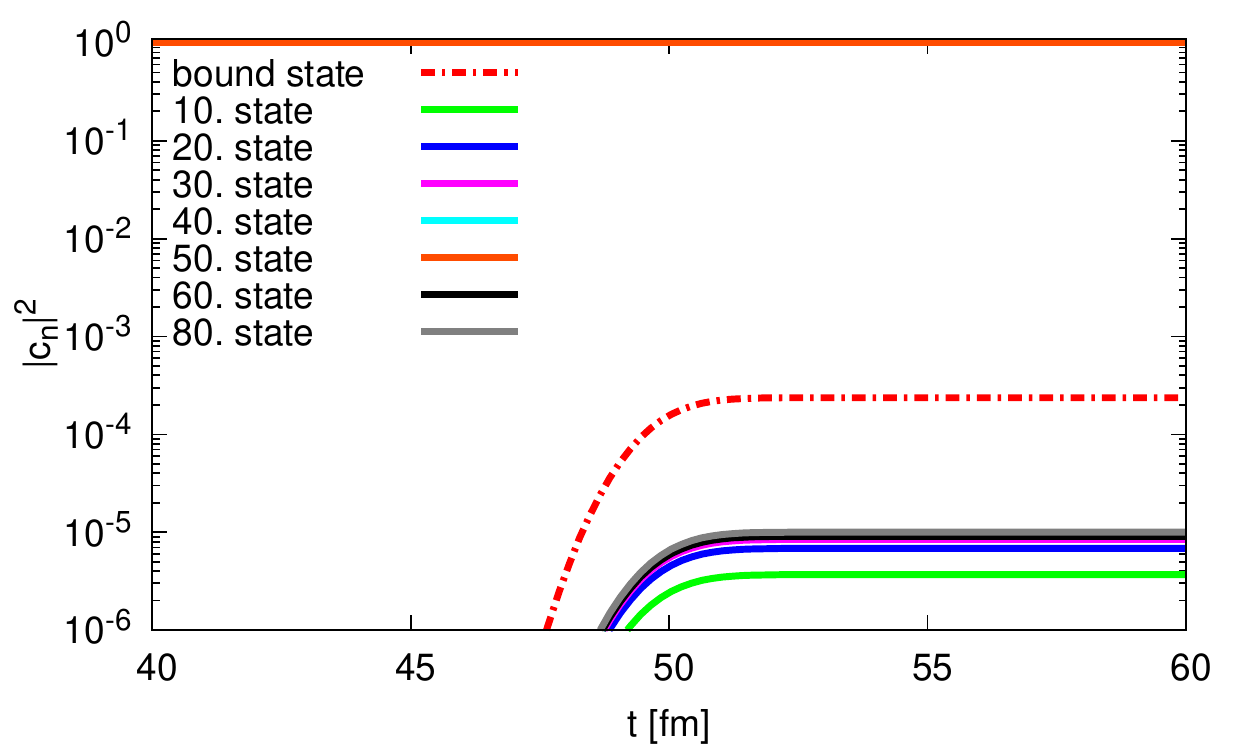}
 	\caption{$|c_n (t)|^2$ with a time dependent potential. Here $|c_{50}(t=0)|=1$ and $|c_{n\neq 50}(t=0)|=0$,  $\sigma_t = 1 $ fm and $\sigma_x = 0.2 a=0.12$ fm, $t_0 = 50$ fm and $V=100$ MeV.}
 	\label{fig:nstate}
 \end{figure}

This set of first-order differential equations can be simplified by the ansatz
 \begin{align}\label{eq22}
 c_j(t) = \tilde{c}_j(t) \ee^{-\text{i}E_jt/\hbar},
 \end{align}
which leads to

\begin{equation}
   \label{eq22.1}
 \text{i} \hbar \dot{ \tilde{c} }_j(t)= \sum_n V_{jn}(t)\exp\left (\text{i}\omega_{jn}t\right) \tilde{c}_n (t),
\end{equation}
where we have defined the transition frequencies
$\omega_{jn}= (E_j-E_n)/\hbar$. 

For the numerical solution of the infinite coupled set of differential
equations \cref{eq22.1} we truncate the expansion by using the first 110
eigenstates only and use a fourth-order Runge-Kutta solver. Then
$\sum_n |c_n(t)|^2 = 1$ for all $t$. To evaluate the accuracy of the
numerical results, we show this conservation of the normalization of the
state, which should hold exactly since in the truncated Hilbert space
the matrix $V_{nm}=\braket{n|\hat{V}| m}$ is Hermitian. As illustrated
in \cref{fig:norm_t} (the parameters are detailed in \cref{sec:POTENTIAL} and \ref{heisenberg})
the norm of the numerically calculated state at $t=100 \; \fm$ deviates
from 1 only by about $10^{-6}$, and proves the high
accuracy of the numerical integration of the coupled set of linear
differential equations (\ref{eq22.1}).

\section{Time dependent Potential and Dynamics of States}
\label{sec:POTENTIAL}
\begin{figure}
	\centering
	\includegraphics[width=\columnwidth]{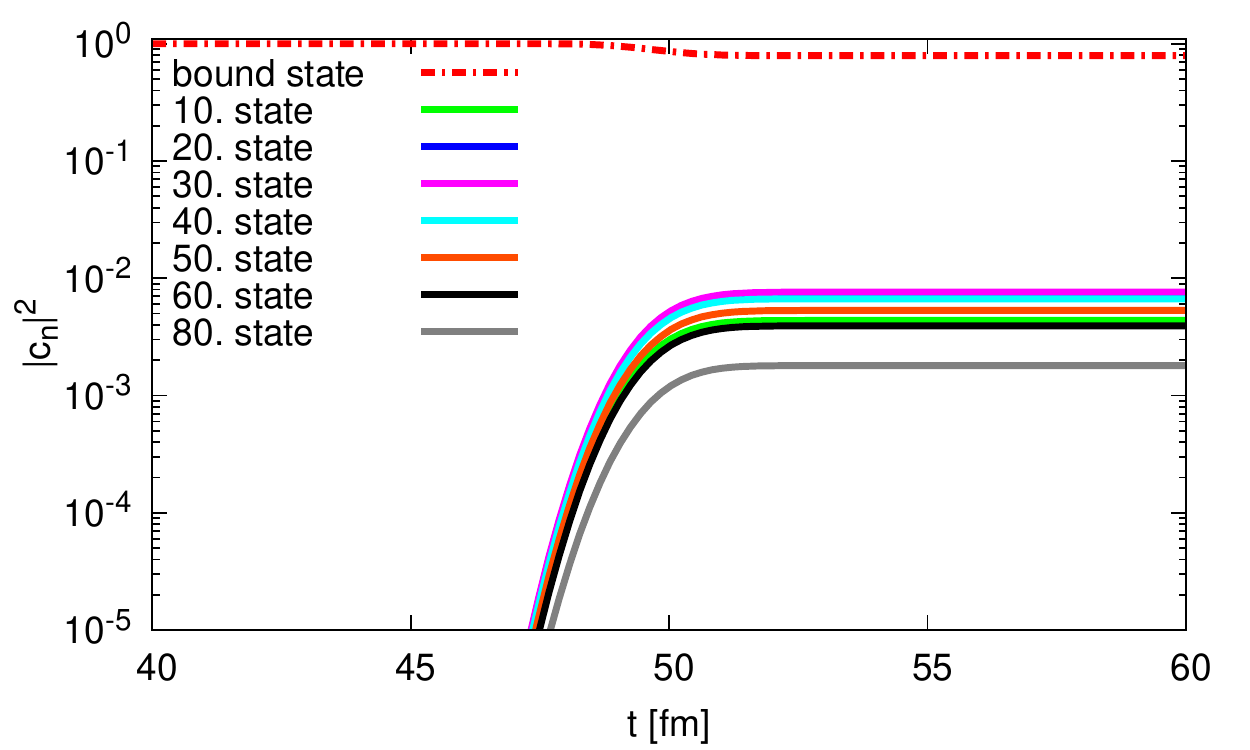}
	\caption{$|c_n (t)|^2$ with a time dependent potential. The bound state is originally populated, $|c_0(t=0)|=1$ and $|c_{n\neq 0}(t=0)|=0$,  $\sigma_t = 1 $ fm and $\sigma_x = 1.2$ fm, $t_0 = 50$ fm and $V=100$MeV.}
	\label{fig:0_x1state}
\end{figure}
\begin{figure}
\centering
\includegraphics[width=\columnwidth]{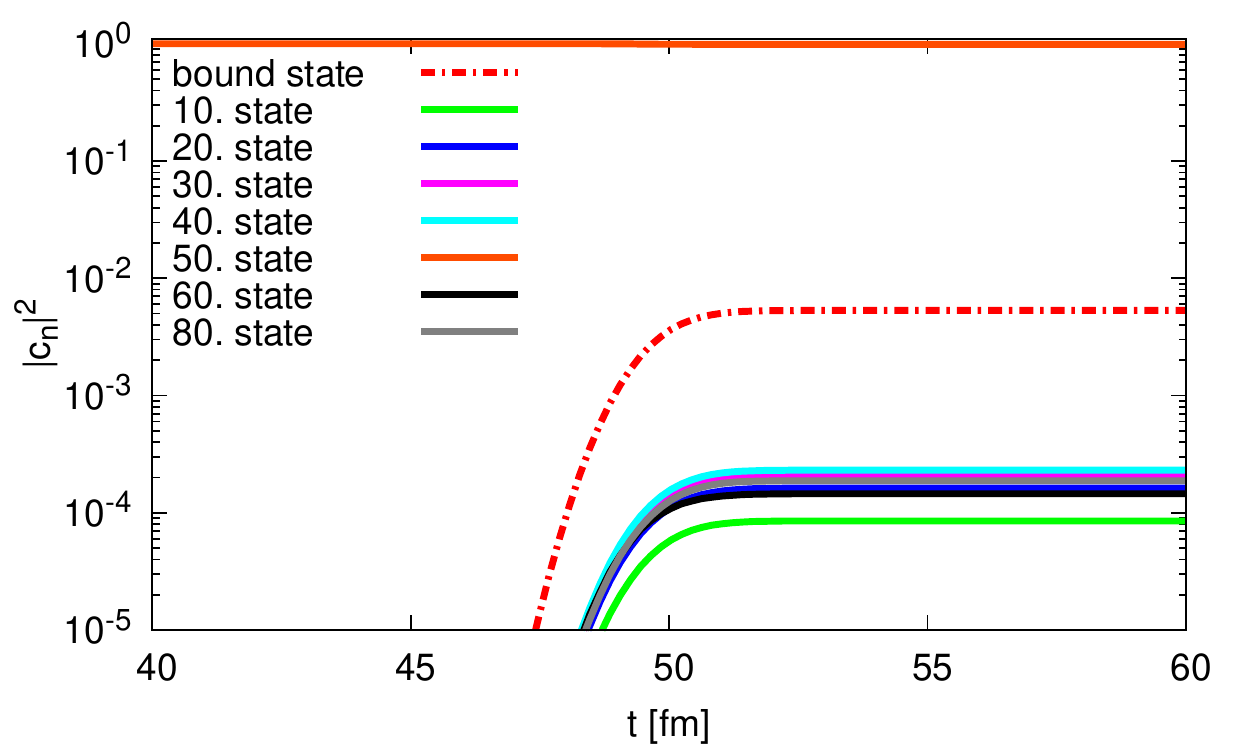}
\caption{$|c_n (t)|^2$ with a time dependent potential. Here $|c_{50}(t=0)|=1$ and $|c_{n\neq 50}(t=0)|=0$,  $\sigma_t = 1 $ fm and $\sigma_x = 1.2$ fm, $t_0 = 50$ fm and $V=100$ MeV.}
\label{fig:n__x1state}
\end{figure}

As already discussed, the  ``deuteron parameters'' for the one-dimensional
square well potential  in a box are given by a potential depth of $V_0 =-18$ MeV, a diameter of $2a = 1.2$ fm and a binding energy of $-2.3$ MeV. This
value of $V_0$ deviates from the value given in \cite{povh14}, as there the
three-dimensional case of a spherical symmetric cavity is considered with $V_0 = -57$ MeV. However, for simplification we discuss the
one-dimensional case here.  In the following calculations we
take a time dependent Gaussian potential, which reads
\begin{equation}
\begin{split}
\label{potential}
V(x,t) = &V \exp\left [- \frac{(x-x_0)^2}{2\sigma_x^2}\right]
\left[\exp\left(- b (t-t_0)^2\right)\right. \\
& \left.+\exp\left(- b (t-t_1)^2\right) + \cdots \right .\\
&\left .+\exp\left(- b (t-t_N)^2\right) \right] ,
\end{split}
\end{equation}
with $b = \frac{1}{2\sigma_t^2}$, where $t_0, t_1,..., t_N $ are the
$N$-times, when ``potential pulses'' interact with the system. We take $x_0 = 0$ as the potential should disturb the system where the bound/ground state is well localized. 
\begin{figure}
	\centering
	\includegraphics[width=\columnwidth]{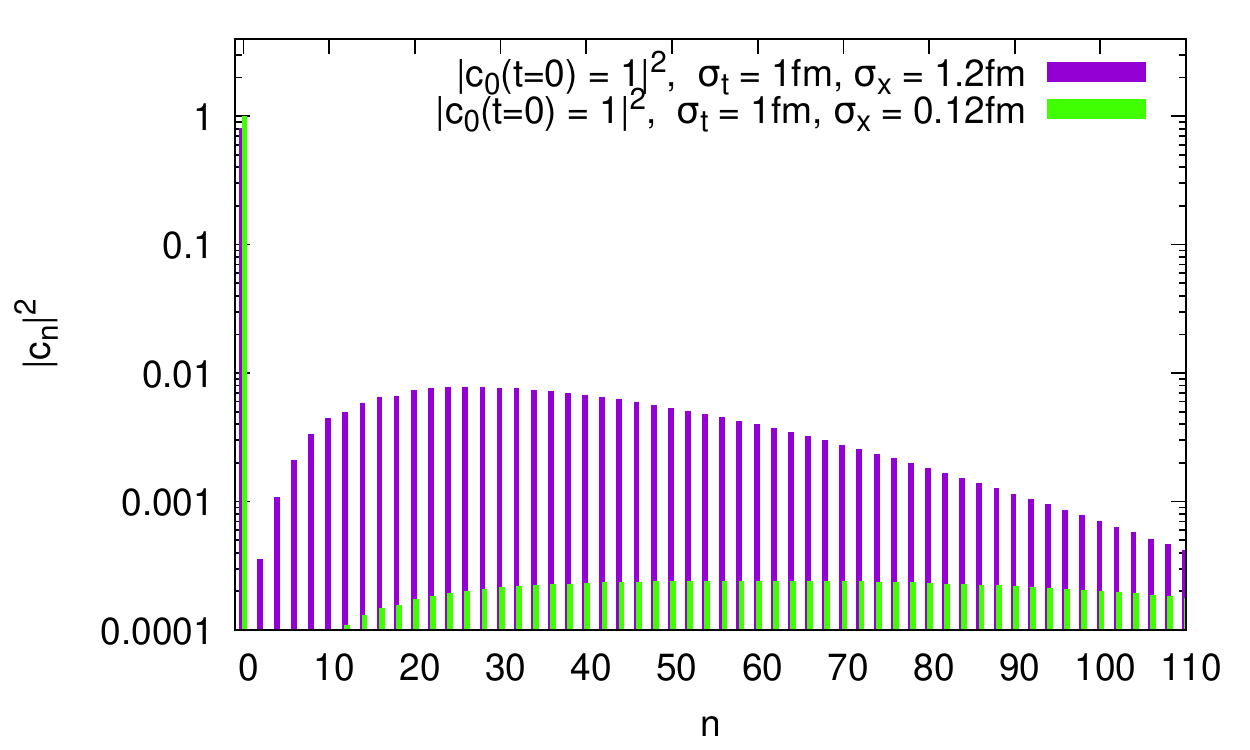}
	\caption{$|c_n (t\gg t_0)|^2$ after the impact of the time dependent potential. Here $|c_{0}(t=0)|=1$ and $|c_{n\neq 0}(t=0)|=0$ for different space width, cf. \cref{fig:0state} and \cref{fig:0_x1state}. $V=100$ MeV.}
	\label{fig:histo}
\end{figure}
\begin{figure}
	\centering
	\includegraphics[width=\columnwidth]{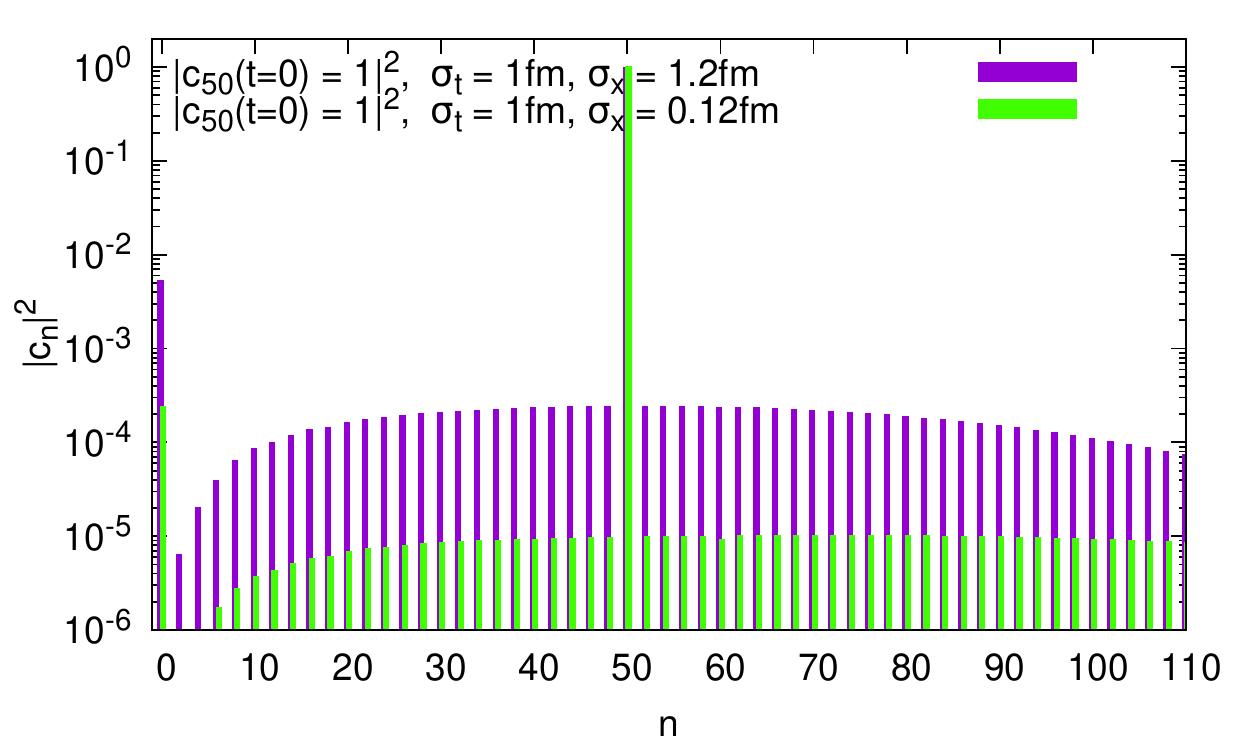}
	\caption{$|c_n (t\gg t_0)|^2$ after the impact of the time dependent potential. Here $|c_{50}(t=0)|=1$ and $|c_{n\neq 50}(t=0)|=0$ for different space width, cf. \cref{fig:nstate} and \cref{fig:n__x1state}. $V=100$ MeV.}
	\label{fig:histo50}
\end{figure}

\begin{figure}
	\centering
	\includegraphics[width=\columnwidth]{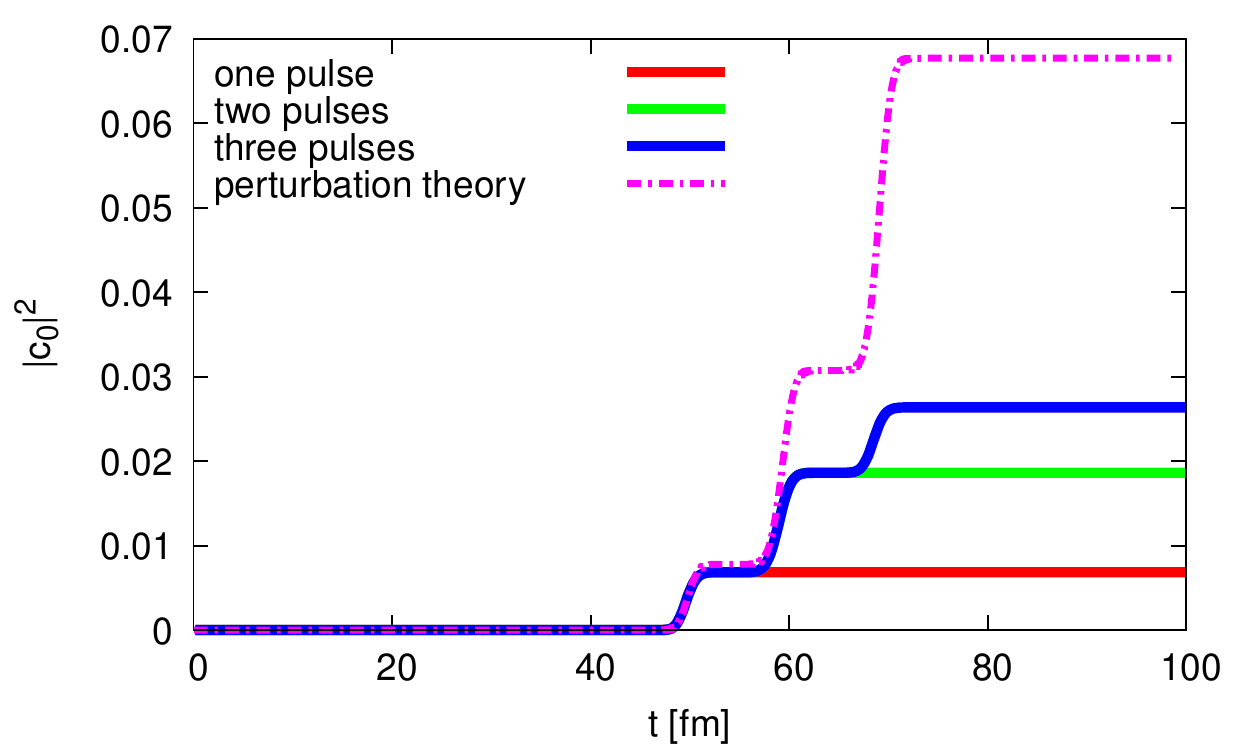}
	\caption{$|c_0 (t)|^2$. Bound state formation after three pulses, with $|c_{20} (t=0)|^2=1$, $\sigma_t=1$ fm, $\sigma_x=1.2$ fm, $V = 100$ MeV. The dashed line represents the same calculation in first order perturbation theory (cf. \cref{sec:pert}).}
	\label{fig:n0}
\end{figure}

At first the situation is given by the interaction of one single pulse.
The parameters, which were chosen here are either $\sigma_x = 2a=1.2$ fm or $\sigma_x = 0.2a=0.12$ fm to
study the impact of the spatial width of the potential and apply various
time durations $\sigma_t$. In \cref{fig:0state} one can see, that if the
bound state, corresponding to the binding energy of a deuteron, $-2.3$
MeV, is originally populated, $\vert c_0(t=0)\vert^2 =1$, and the system
interacts with a time dependent potential at some time, here for
$t_0 = 50$ fm and $\sigma_t = 1$ fm, the excited states are populated
right at the arrival of the pulse, and the originally prepared bound
state decreases correspondingly. For $\sigma_x = 0.2a$ the bound state
is depopulated by about $4\%$, and the system has gained a small energy increase due
to the interaction, cf. \cref{fig:0state}. On the other hand, if
$\sigma_x$ is larger, here $2a$, then the originally populated bound
state is depopulated by about $30\%$, also the mean population of the
other states grows by orders of magnitude, cf. \cref{fig:0_x1state}.

One can observe this behavior also if the 50th state is populated
originally and then decays after the impact of the potential,
cf. \cref{fig:nstate} and \cref{fig:n__x1state}. In both cases,
$\sigma_x = 0.2a$ and $\sigma_x = 2a$, $|c_{50} (t\gg t_0)|^2$ ($E_{50} \approx 26$\;MeV, cf. \cref{fig:energy})
decreases slightly, while the bound state reacts strongly and also
almost instantaneously on the time scale $\sigma_t$.  Due to the large
energy gap the bound state populates by more than an order of magnitude stronger than the other excited states.
In \cref{fig:histo} one can see the distribution of populated states if
the bound state is originally prepared and $\sigma_x$ is varied between
$\sigma_x = 0.2a$ and $\sigma_x =2a$. All the other parameters in
\cref{potential} are kept fixed at $\sigma_t = 1\,\text{fm}$ and
$V=100 \, \text{MeV}$. One finds a small decrease of the bound state,
$n=0$, and an increase of all the other states after the impact of the
potential.

\begin{figure}
	\centering
	\includegraphics[width=\columnwidth]{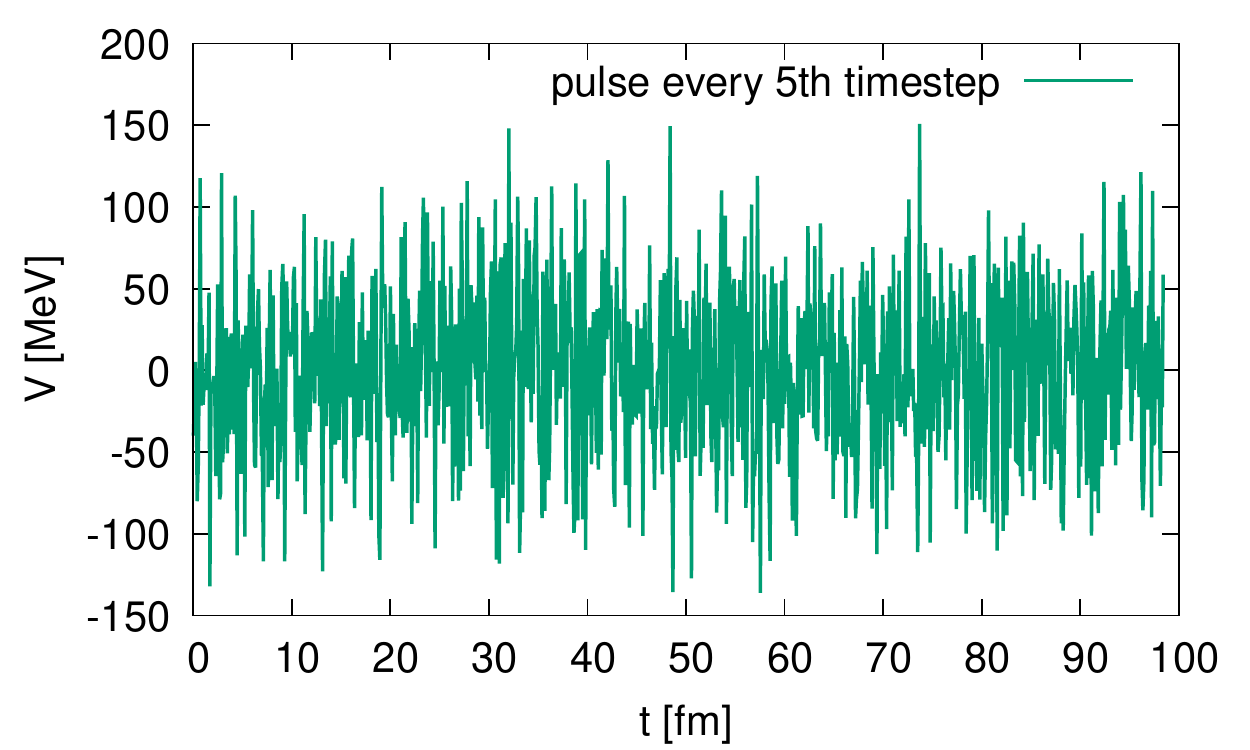}
	\caption{Stochastic potential with 1000 equidistant timesteps
          and Gaussian distributed potential strength.}
	\label{fig:random}
\end{figure}
\begin{figure}
	\centering
	\includegraphics[width=\columnwidth]{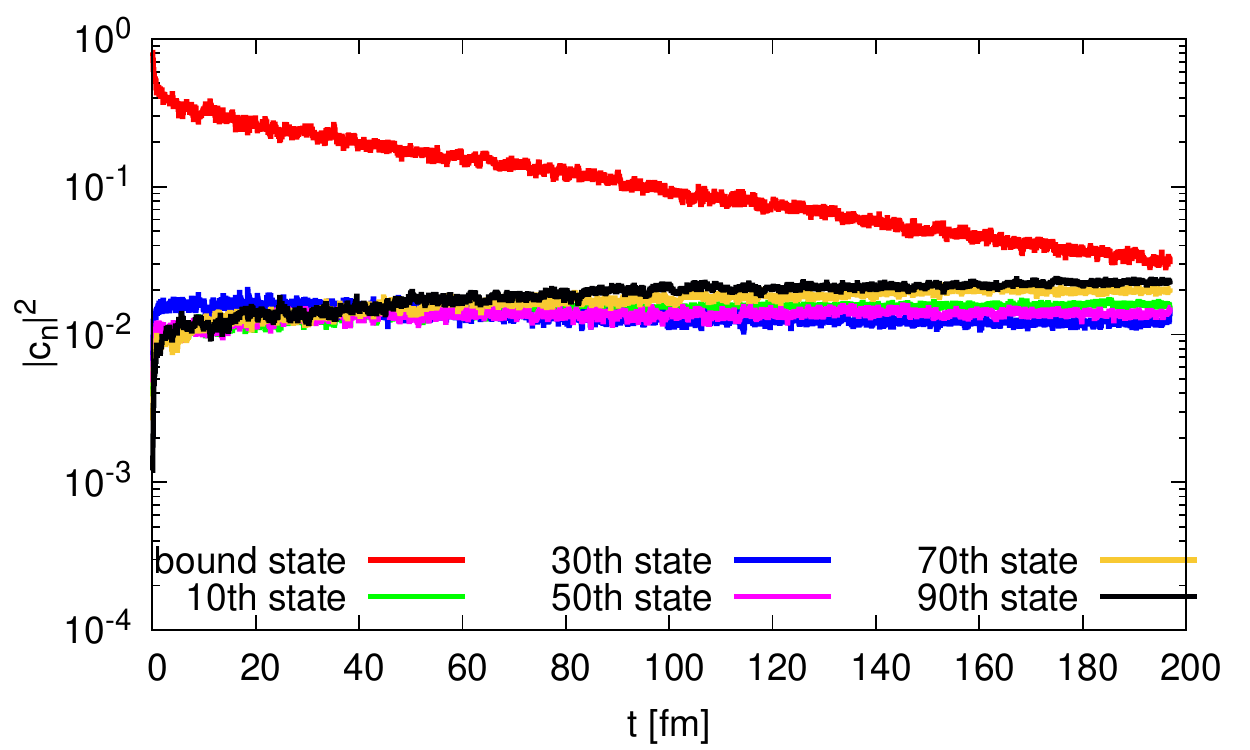}
	\caption{Averaged $|c_n(t)|^2$ of a Gaussian distributed
          stochastic potential with 2000 pulses with initial conditions
          $|c_{0}(t=0)|^2 = 1$ and $|c_{n\neq 0}(t=0)|^2 = 0$. Here
          $\sigma_x = 1.2$ fm and the potential given in \cref{pot}.}
	\label{fig:stochastic}
\end{figure}
\begin{figure}
	\centering
	\includegraphics[width=\columnwidth]{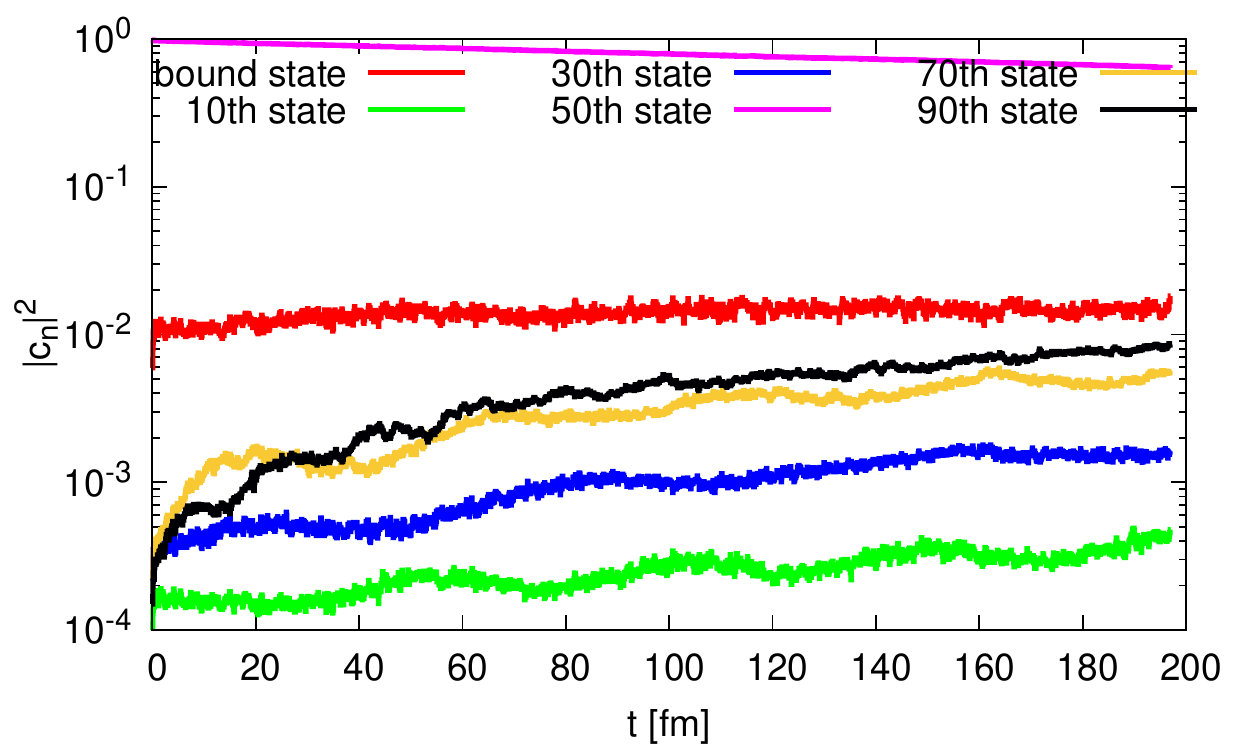}
	\caption{Averaged $|c_n(t)|^2$ of a Gaussian distributed
          stochastic potential with 2000 pulses with initial conditions
          $|c_{50}(t=0)|^2 = 1$ and $|c_{n\neq 50}(t=0)|^2 = 0$. Here
          $\sigma_x = 1.2$ fm and the potential given in \cref{pot}.}
	\label{fig:n50}
\end{figure}
It is important to mention, that increasing the spatial width of the
pulse does not lead to a linear increase in the state formation, but
also lead to a different shape of the distribution in the states. One
can see in \cref{fig:histo}, that a larger $\sigma_x$ narrows the
distribution, such, that they are Breit-Wigner-like distributed around
the 25th state, which corresponds to an energy of
$\sim 3 \, \text{MeV}$, cf. \cref{fig:energy}.  If the
50th state is populated originally, then after the impact of the pulse,
$\vert c_{50}(t)\vert^2$ decreases slightly, as already mentioned. The
bound state forms dominantly, compared to the other states. The most
important difference is, that the states populate in a broad
distribution, whose broadness is not dominantly determined by $\sigma_x$,
cf. \cref{fig:histo50}. Of course a wider pulse leads to a stronger
population of the states.

 We will continue the discussion of the 
results of \cref{fig:histo} and  \cref{fig:histo50} in \cref{heisenberg}, where the impact of
the interaction time $\sigma_t$ on the distribution of states will be investigated in more depth, 
in order to understand the interplay between the energy distribution and the pulse duration 
in terms of Heisenberg's uncertainty relation of energy and time.
The impact of
a longer and a shorter pulse and taking $\sigma_x=1.2$ fm
refers to a particle, which interacts with a deuteron, having a similar
size in the interaction channel.

Following \cref{potential}, one also can easily increase $N$, the number of pulses, for example $N = 2$ and $N = 3$,
cf. \cref{fig:n0}. One can see, that after each pulse, the bound state will be increasingly populated.
 The other states will populate similarly in time (with other amplitudes) and the
20th state will depopulate accordingly. In addition, the results (purple dashed line) of
first order perturbation theory for this particular situation is also depicted and will be
discussed briefly in \cref{sec:pert}.

 \begin{figure}
 	\centering
 	\includegraphics[width=\columnwidth]{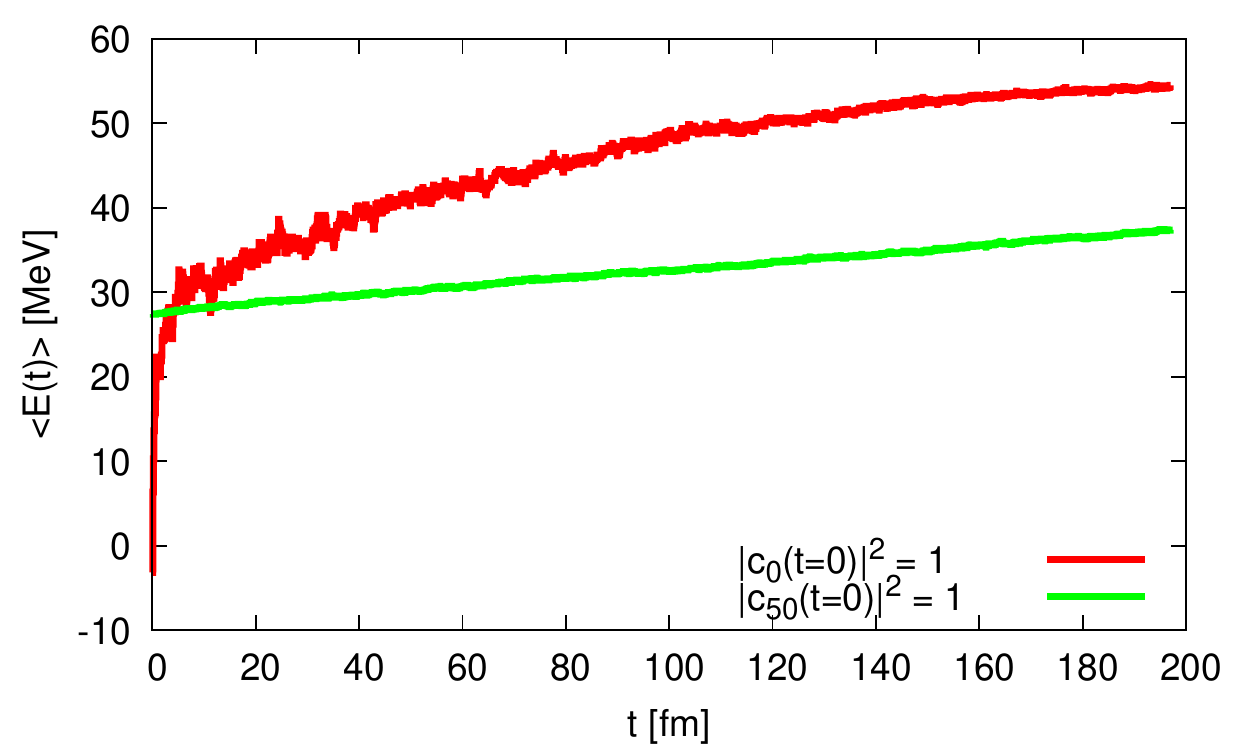}
 	\caption{Mean energy, cf. \cref{energyform}, of a Gaussian
          distributed stochastic potential with 2000 pulses.}
 	\label{fig:mean_energy}
 \end{figure}
\section{Randomized potential pulses}\label{sec:random}

To briefly study the behavior of a system, that interacts all the time with a stochastic
potential, such as a random noise, we increase the number of pulses $N$ to an order of
$\mathcal{O}(2000)$, where $t_{i+1} = t_i + \Delta t$, with a Gaussian
distributed amplitude $V_{\xi}$ with  $\braket{V_{\xi}} = 0$ and
$\sigma_V = 50$ MeV,
\begin{equation}
\begin{split}
\label{pot}
V(x,t) = &\sum_{j=1}^N V_\xi \exp \left
  (-\frac{x^2}{2\sigma_x^2} \right) \\
&\left[\Theta(t-j n\Delta t) - \Theta(t-j (n+1)\Delta t) \right].
\end{split}
\end{equation}
Furthermore, the parameters are taken as $\Delta t = 0.02 \,\fm$ and
$\sigma_x = 1.2 \,\fm$, and the real value $n$ is set to be $n=5$, which
means, that numerically every 5th time step a random pulse occurs. This
corresponds to the interaction of the nuclei with random ``bath
particles'', which leads to the formation and dissociation of bound
states or other quantum states as described above. Generating this
statistical potential leads to the potential profile illustrated in
\cref{fig:random}.

In the following we take ensemble averages of the time evolution over
these random potentials. Figs.\ \ref{fig:stochastic} and \ref{fig:n50}
show the evolution of states under the influence of a potential with
2000 pulses, employing two different initial conditions: (a) the bound
state originally populated and (b) the 50th state originally populated;
the results are averaged over 200 Monte-Carlo realizations of the random
process, $V(x,t)$. It turns out that the initially prepared state is
depopulated continuously and the other states get correspondingly
occupied. As illustrated in \cref{fig:n50} the bound state forms on a
very small timescale, i.e., rather instantly.
The bound state reacts fastest to the impact of the pulses,
cf. \cref{fig:n50}, red line, since the state of smallest and distinct
energy is preferably populated.

As illustrated in \cref{fig:mean_energy}, the mean energy of the system,
\begin{equation}\label{energyform}
\braket{E(t)} = \frac{\sum_n E_n \vert c_n(t) \vert^2}{\sum_n  \vert
  c_n(t) \vert^2} = \sum_n E_n \vert c_n(t) \vert^2,
\end{equation}
starts at $t=0$ with the $n$-th energy eigenvalue of the initialized
state and then continuously increases due to the time-dependent external
potential. This behavior as found stands in contrast to a potential
equilibration of the system, that should occur for
$t \rightarrow \infty$, if we consider the system as coupled to a
potential heat bath of random particles, or interactions. In principle
energy dissipation has to be incorporated as an additional ingredient
for the propagation of the particle in the system, in analogy to the
classical Langevin equation,
\begin{align}
 \dot{p} = -\gamma p + f(t).
\end{align}
Dissipation and fluctuation are intimately related. The further
development of a consistent quantum-theoretical description for the
formation of bound states in open quantum systems is an intriguing
question, and is relevant in particular for the microscopic
understanding of the production of light nuclei
\cite{DANIELEWICZ1991712} or heavy quarkonia states in relativistic
heavy ion collisions \cite{Katz_2016,Blaizot_2018,Rothkopf_2020}. Powerful
non-equilibrium quantum formalisms to evaluate the equilibration of an
open quantum system, including dissipation, are given by the
Kadanoff-Baym equations \cite{DANIELEWICZ1984239,GREINER1998328}
or by the Caldeira-Leggett master equation \cite{CALDEIRA1983587}, both
of which describe the time evolution of the system in terms of (reduced)
density matrices or Green's functions.

\begin{figure}
	\centering
	\includegraphics[width=\columnwidth]{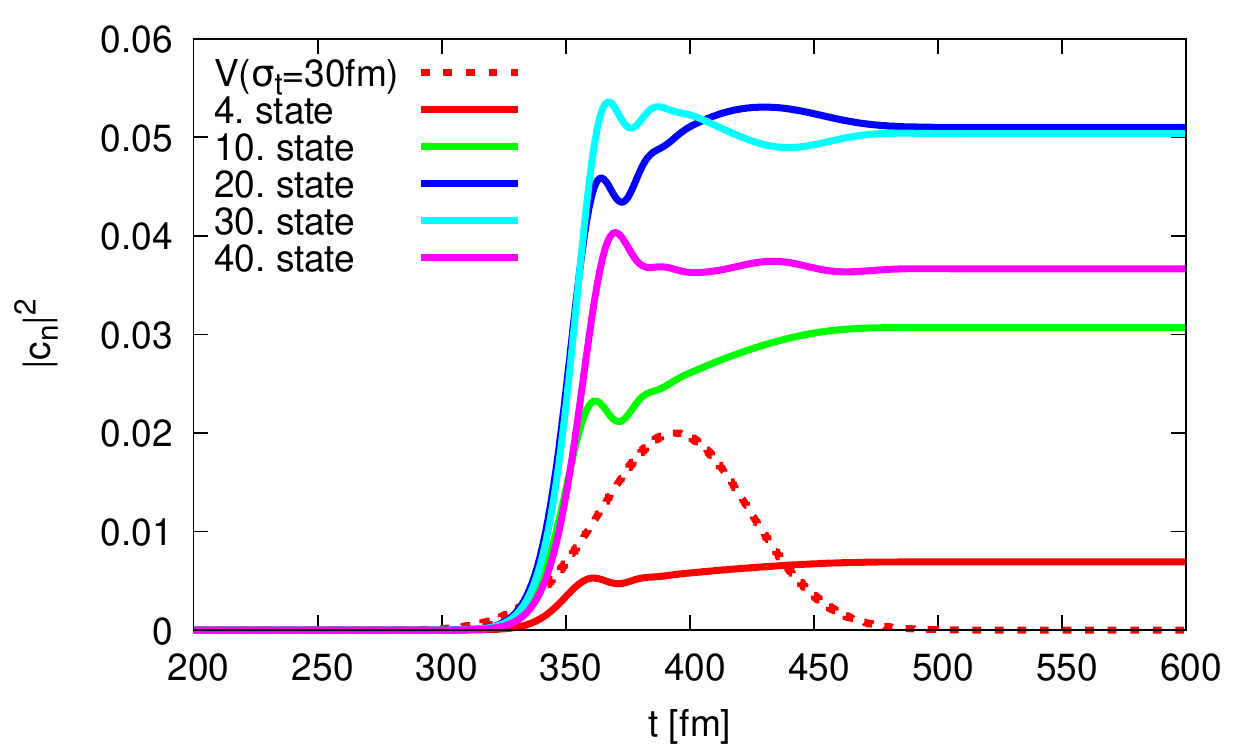}
	\caption{Non-perturbative behavior of the system, reacting on a long pulse, here $\sigma_t=30$ fm, $\sigma_x =1.2$ fm and $V=100$MeV, centred at $t_0=400$ fm and $\vert c_0(t=0)\vert^2=1$. Green dashed line indicates the shape of the pulse.}
	\label{fig:longpulse}
\end{figure}
\begin{figure}
	\centering
	\includegraphics[width=\columnwidth]{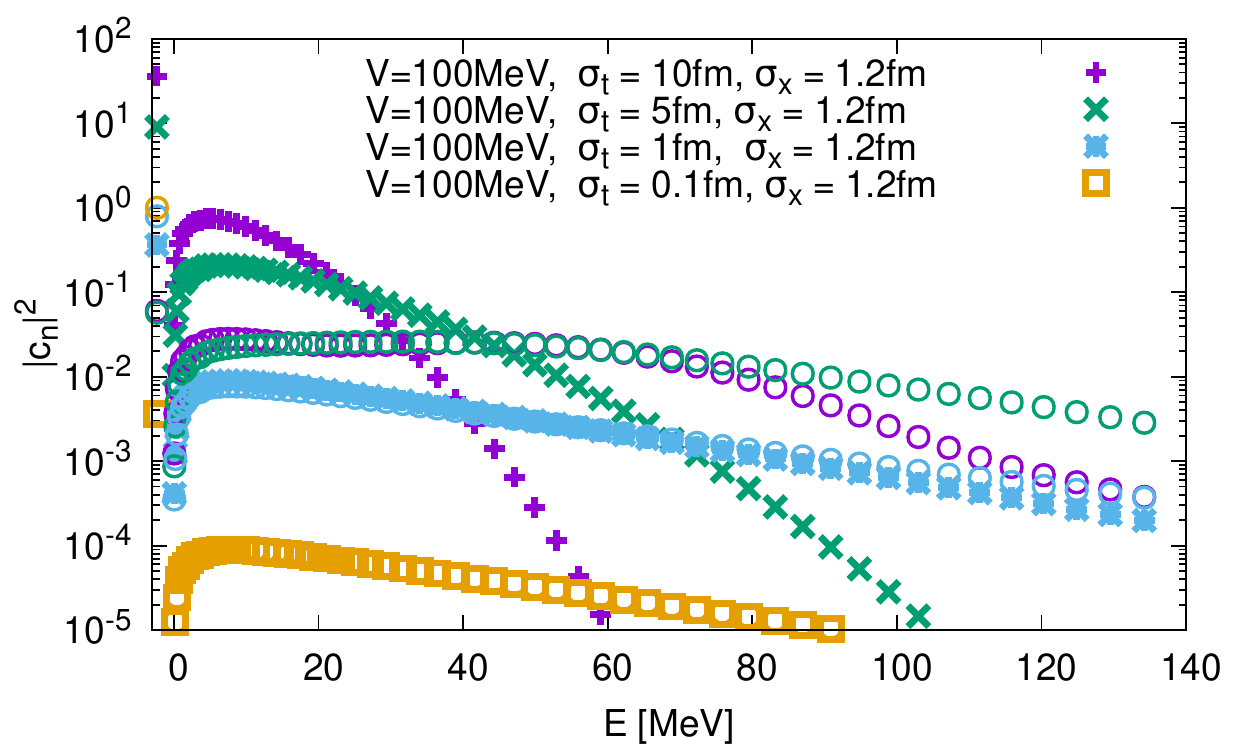}
	\caption{First order perturbation theory for different $V$, $\sigma_t$ and $\sigma_x$. The circles represent the exact numerical solution corresponding to the same color in perturbation theory.}
	\label{fig:pert}
\end{figure}
\begin{figure}
	\centering
	\includegraphics[width=\columnwidth]{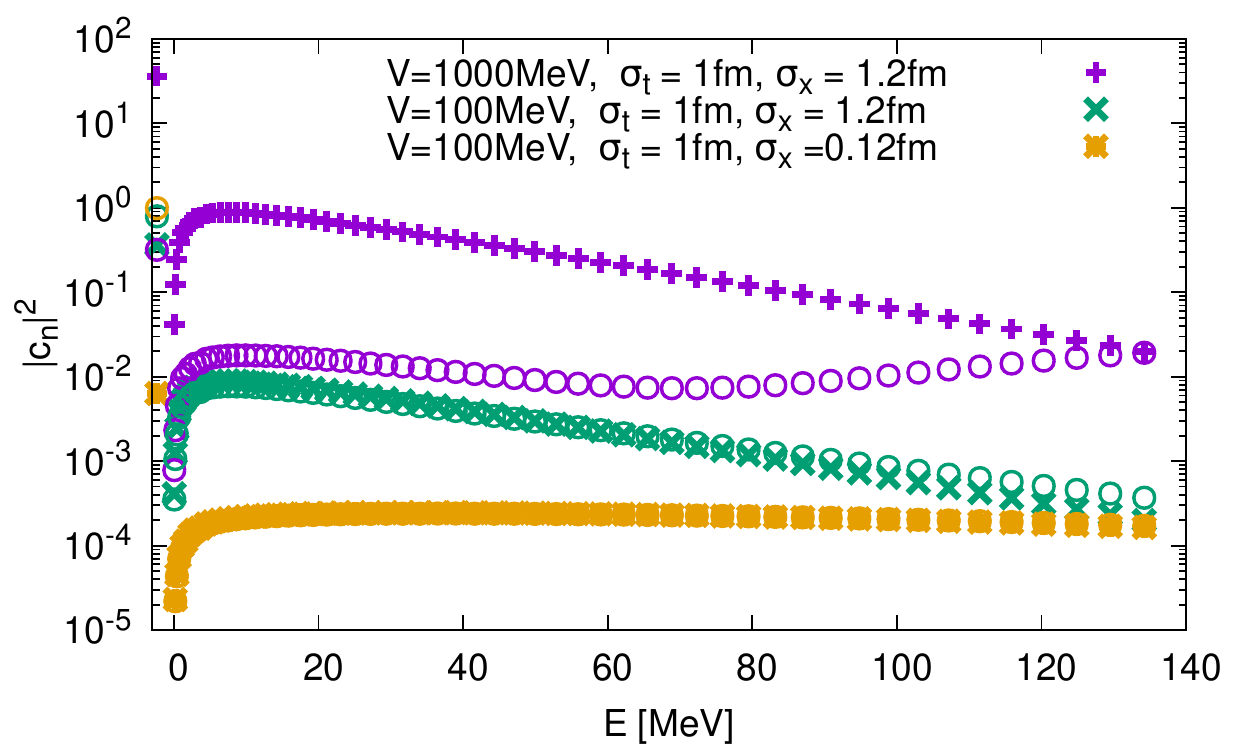}
	\caption{First order perturbation theory for different $V$ and $\sigma_x$. The circles represent the exact numerical solution corresponding to the same color in perturbation theory.}
	\label{fig:pertV}
\end{figure}


\begin{figure}
	\centering
	\includegraphics[width=\columnwidth]{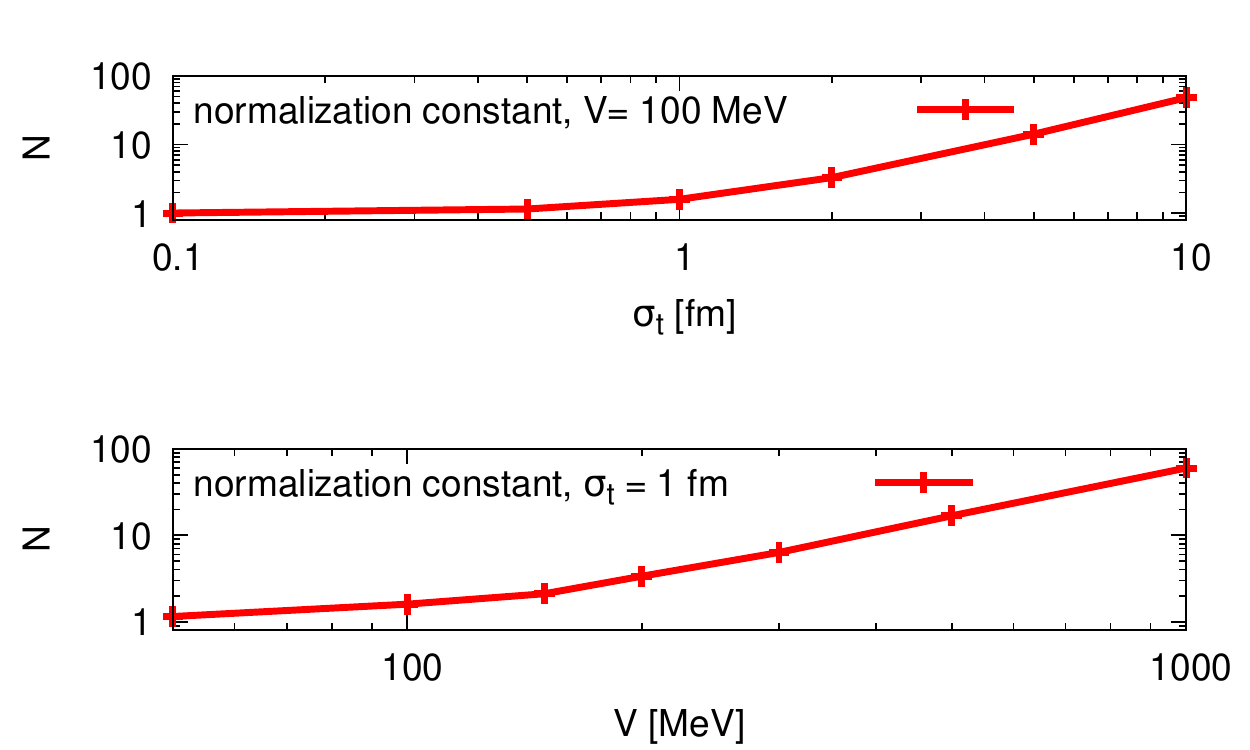}
	\caption{$N$ in first order perturbation theory for different $V$, $\sigma_t$ at $\sigma_x=1.2$ fm. }
	\label{fig:norm}
\end{figure}

\section{Perturbation theory}
\label{sec:pert}

In this Sec.\ we also want to study the applicability of perturbation theory, i.e.,
we validate Fermi's golden rule, i.e., first order perturbation theory,
which allows the description of the transition amplitude from an initial
state $i$ to a final state $f$ via $|c_n(t)|^2$,

\begin{align}\label{first}
c_{f}^{(1)}(t) &= \delta_{fi} - \frac{\text{i}}{\hbar} \int_{t_0}^t \text{d}t' V_{fi} (t'),
\end{align}
where $V_{fi}(t') = V(t') \exp(-\text{i}\omega_{fi} t')$
\cite{Landau1981Quantum}.  To set the stage, in \cref{fig:longpulse} 
the highly non-perturbative behavior of a long pulse is clearly demonstrated.


Intuitively and regarding \cref{first} first order perturbation theory
should be valid if
$ \frac{1}{\hbar^2} \vert \int \text{d}t V_{fi}\vert^2\ll
1$. Furthermore, (cf. \cite{normalization}) first order perturbation
theory is not normalized. Therefore it is a reasonable method to
evaluate the applicability of perturbation theory by considering the
``normalization constant'', $N = 1/\sqrt{\sum_n |c_\text{n,
    pert}|^2}$. In the case of \cref{fig:longpulse}, $N\approx 74$,
which is much larger than one and proves, that here perturbation theory
is not applicable.

To study the validity of perturbation theory for different model
parameters, in \cref{fig:pert} $\vert c_n(t\gg t_0)\vert^2$
different pulse lengths $\sigma_t$ are shown. In \cref{fig:pertV} the
potential strengths are set to either $V=100$ MeV or $V=1000$ MeV and also
the widths are either $\sigma_x=0.12 \,\fm$ or $\sigma_x=1.2\;\fm$, while $\sigma_t$ is set constantly at 1 fm and
first-order perturbation theory is compared to the exact numerical results in the
case of one pulse at $t_0=50\,\fm$. The circles of the same color refer
to the same parameters as the perturbative ones but represent the exact
results solving the time-dependent Schr\"odinger equation. In
\cref{fig:norm} the normalization constant, $N$, is shown for
different values of $\sigma_t$ while keeping $V=100 \, \MeV$ constant,
and for different values of $V$ while keeping $\sigma_t=1 \,\fm$
constant.

As shown in \cref{fig:pert}, for $\sigma_t = 1$ fm and $\sigma_t = 0.1$
fm for a potential of $V=100$ MeV and $\sigma_x=1.2$ fm perturbation
theory and the numerical results are still applicable and thus in reasonable
agreement. If one increases $\sigma_t$, such that $\sigma_t=5$ fm or
$\sigma_t = 10$ fm the perturbative results
deviate from the numerical ones, not only by a factor but also in the
shape of the distribution. For $V = 1000$ MeV instead,
cf. \cref{fig:pertV}, all perturbative results lie about two orders of
magnitude above the exact numerical results.

As expected, the perturbative results for $V=1000$ MeV deviate from the
ones for $V=100$ MeV by a factor of 100, cf. \cref{first}. Furthermore
the shape of the distribution changes, because for the numerical results
the $\vert c_n\vert^2$ increase for higher energies. This is due to the
fact, that $V=1000$ MeV exceeds the highest energy of
$\approx 140 \; \MeV$ taken into account in the truncated Hilbert
space. On the other hand, the norm is conserved due to the unitarity of
the time evolution in the truncated Hilbert space,
$\sum_n\vert c_n\vert^2=1$, which leads to a higher occupation of the
higher states.  This of cause does not affect the calculation in
perturbation theory.


For $V=100$ MeV, $\sigma_t=1 \,\fm$ but $\sigma_x = 0.12 \,\fm$
perturbation theory agrees satisfactorily with the exact numerical
results.  Considering a typical pulse duration of an interacting
particle having a similar size as a deuteron ($\sigma_x=1.2 \;\fm$), of
about $\sigma_t=1 \;\fm$ and $50$-$150 \;\MeV$, thus agrees reasonably
well with the exact calculation. As depicted in \cref{fig:norm},
increasing the potential, $V$, and/or the duration of the pulses,
$\sigma_t$, leads to an increase of $N$ and therefore the failure of
perturbation theory. E.g., for $\sigma_t \approx 5$ fm $N$ is already
about $10$ times larger than for $\sigma_t = 1$ fm. For larger
potentials, the normalization diverges more or less quadratically. In a
range of $\sigma_t =1$ fm and $V=100$ MeV, $N$ is about 1, indicating
that perturbation theory is a good approximation for this parameter
setting.

Therefore one can conclude, that perturbation theory is applicable for
weak potentials in comparison to the considered energy range of the
system and for short interaction times of about 1\,fm. Furthermore, one
can see, that an increase of either parameter leads to an increase of
the error of the perturbative results, which is due to the fact, that
$\frac{1}{\hbar^2}\vert \int \text{d}t V_{fi}\vert^2\ll 1$ should hold as an
applicability criterion for first order perturbation theory.
Correspondingly, if one increases the number of pulses,
cf. \cref{fig:n0}, where the number of three pulses has been considered, the
perturbative calculation provides significantly different results than
the numerical calculation due to the fact, that in the integral of
\cref{first} the contributions from the subsequent pulses just add up,
while in the full calculation all states get populated already after the
first pulse considerably, which is not taken into account in
\cref{first}. In order to improve the applicability of perturbation theory, one has to modify the ansatz to obtain correct
rate equations in form of a (norm-conserving) master
equation, which takes into account gain and loss of the states after
every single pulse. This is the legitimation of a (quantum) kinetic master equation or Boltzmann-type equation.
  \begin{figure}
	\centering
	\includegraphics[width=\columnwidth]{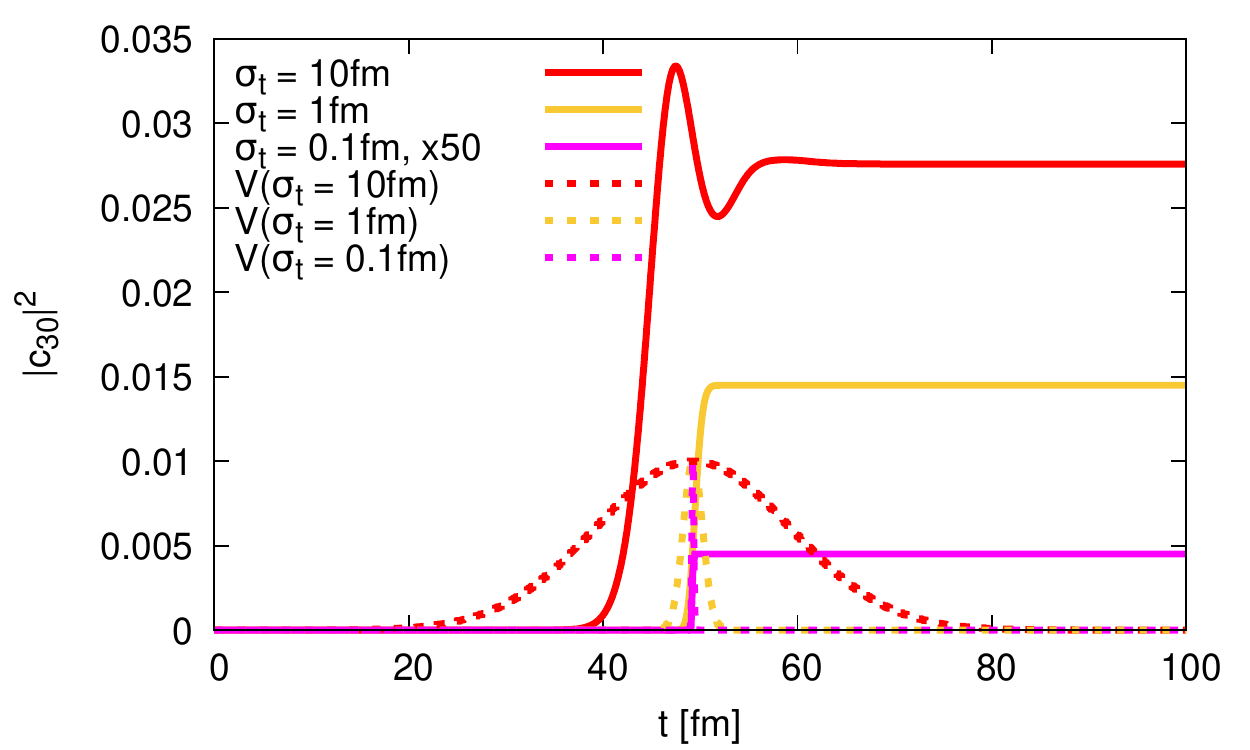}
	\caption{Formation of states during a pulse with $\sigma_t = 10$
          fm, $\sigma_t = 1$ fm and $\sigma_t = 0.01$ fm (here
          $\vert c_{30}(t, \sigma_t = 0.1\text{ fm} )\vert^2$ multiplied
          by a factor of 50, for visualization), $\sigma_x = 1.2$fm and
          $V=100$MeV, $\vert c_0(t=0)\vert^2=1$. }
	\label{fig:lengths}
\end{figure}
\begin{figure}
	\centering
	\includegraphics[width=\columnwidth]{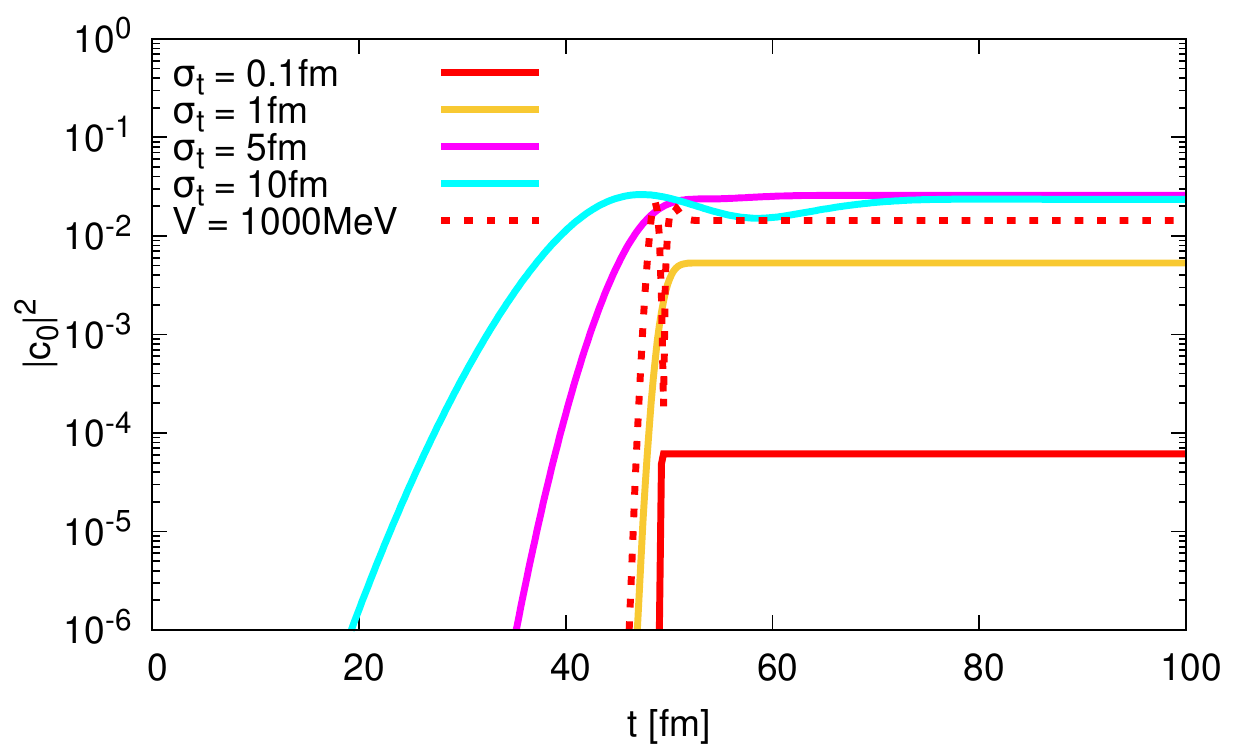}
	\caption{Bound state formation of $\vert c_{50}(t=0) \vert^2 =
          1$ for different time length (solid lines) at $V=100$ MeV and
          a comparison of a strong potential of $1000$ MeV with
          $\sigma_t = 1$ fm (dashed line), one pulse.}
	\label{fig:bs_form}
\end{figure}

\section{Heisenberg's energy-time uncertainty relation}
\label{heisenberg}

A seemingly straight-forward expectation is that the formation of states,
especially bound states, underlies the Heisenberg's uncertainty relation
in energy and time. A disturbance of the system caused by an interacting
particle, e.g. a collision or here the time-dependent potential, leads
to a reaction of the system which then rearranges its energy
distribution over the available states. One possible idea is, that this
rearrangement underlies the uncertainty relation in energy and time
such, that the system needs some time $\Delta t$, which is the
difference between the potential impact and the reaction of the system,
which then is dependent on the energy,
\begin{align}
\tau_{\text{f}} \sim \frac{\hbar}{E_{\text{D}}} \quad \text{or} \quad
  \tau_{\text{f}} \sim \frac{\hbar}{E_{\text{B}}},
\end{align}
where $\tau_{\text{f}}$ is then the ``formation time'' of for example a deuteron,
and $E_{\text{D}}$ the energy difference of a certain state before and after the
interaction or even simply the binding energy, $E_{\text{B}}$.

In contradiction to this possible straight-forward idea, as demonstrated
in \cref{fig:lengths} and \cref{fig:bs_form}, the states form
immediately, independently of the pulse duration (cf. also
\cref{fig:0state}-\cref{fig:n__x1state}). Especially in
\cref{fig:lengths} we decrease the pulse duration to only $0.1$ fm (pink
line) and find, that still the state reacts immediately to the pulse.

Therefore, as illustrated by these results, Heisenberg's uncertainty
relation in energy and time should be understood in a different way. It
is more suitable to talk about the population time instead of the
formation time, because we want to point out, that the formation time is
equivalent to the interaction time of the potential, and therefore the
term is misleading. This picture is also in good agreement with the
interpretation of the energy-time uncertainty relation given in
\cite{Landau1981Quantum} and \cite{messiah99}, which is motivated by
considering the transition probability from the energy eigenstate
$\psi_i$ to the energy eigenstate $\psi_f$ of the unperturbed system,
due to the external potential. In first-order perturbation theory the
transition amplitude is given by \cref{first}, which reads, rewritten
for our present case of a potential (\ref{potential}) representing only
one pulse
 \begin{equation}
\begin{split}
\label{c_n}
 c_f(t) =  -\text{i}\frac{V_{fi}}{\hbar} \int_{-\infty}^t \text{d}t'
 & \exp\left(-\frac{(t'-t_0)^2}{2\sigma_t^2}\right) \\
&\exp(\text{i}
 \omega_{fi} t')
\end{split}
 \end{equation}
where $ \omega_{fi} = \Delta E/\hbar$ and
\begin{equation}
V_{fi}= \int_{\R} \dd x V \psi_f^{*}(x) \exp \left [-\frac{(x-x_0)^2}{2
    \sigma_x^2} \right] \psi_i(x).
\end{equation}



In the limit $t \rightarrow \infty$ the probability for a transition of
the particle from an energy eigenstate $\psi_i$ through the perturbation
of duration $\sigma_t$, adiabatically switched on and off via the
Gaussian time dependence, reads
\begin{equation}
\begin{split}
S_{fi}&=\lim_{t \rightarrow \infty} c_f(t) \\
&= -\ii \sqrt{2\pi} \frac{V_{fi}}{\hbar} \sigma_t \exp\left[-\frac{1}{2} \sigma_t^2  \omega_{fi}^2\right],
\end{split}
\end{equation}
and therefore the transition probability is
\begin{align}
\label{gauss}
\vert S_{fi} \vert^2 = 2\pi \frac{\vert V_{fi}\vert^2
  \sigma_t^2}{\hbar^2 }\exp\left[-\sigma_t^2  \omega_{fi}^2\right].
\end{align}
This implies that the smaller $\Delta t \simeq \sigma_t$, the broader is
the distribution of the observed changes of energy. 

In App.\ \ref{Mandelstamm-Tamm} we give a more general, non-perturbative
interpretation of the time-energy uncertainty relation, related to the
accuracy of ``time and energy measurements''. This is in accordance with
\cite{Landau1981Quantum}, where it is pointed out, that the uncertainty
relation in energy and time can not be interpreted as a uncertainty of
measurement in energy at a certain time, as the uncertainty relation in
position and momentum, but is the difference of energies, that are
measured at two different times.

\Cref{fig:lengths} illustrates, that the system reacts immediately to
the pulse. The dashed lines indicate the pulse, and a randomly picked
state reacts immediately to the potential, independently of the pulse
duration. If one decreases $\sigma_t$ even more ($\sigma_t=0.1$ fm),
then the states populate still immediately with the appearance of the
potential. This is valid also for a system, where $n=50$ is originally
prepared and the bound state is populated due to the perturbation, as
can be seen in \cref{fig:bs_form}. Also the occupation of the states
stays constant immediately after the perturbation is switched off,
cf. \cref{fig:lengths} and \cref{fig:bs_form}. As shown in
\cref{fig:lengths}, if one applies a very short pulse, dashed pink line
(potential) and pink line (state), where the pulse is
$\sigma_t = 0.1\;\fm$, the states get populated immediately, as suggested
above. In \cref{fig:norm_t} we show the conservation of the norm during
the time evolution for $\sigma_t = 1$ fm and $\sigma_t = 10$ fm, which
demonstrates the high numerical accuracy of the calculation. In
\cref{fig:bs_form} one can also see, that a stronger potential, here
$1000$ MeV, does not automatically lead to a stronger increase in the
states, but leads to oscillations during the pulse, which shows again,
that first-order perturbation theory is not applicable in this case.
   \begin{figure}
  	\centering
  	\includegraphics[width=\columnwidth]{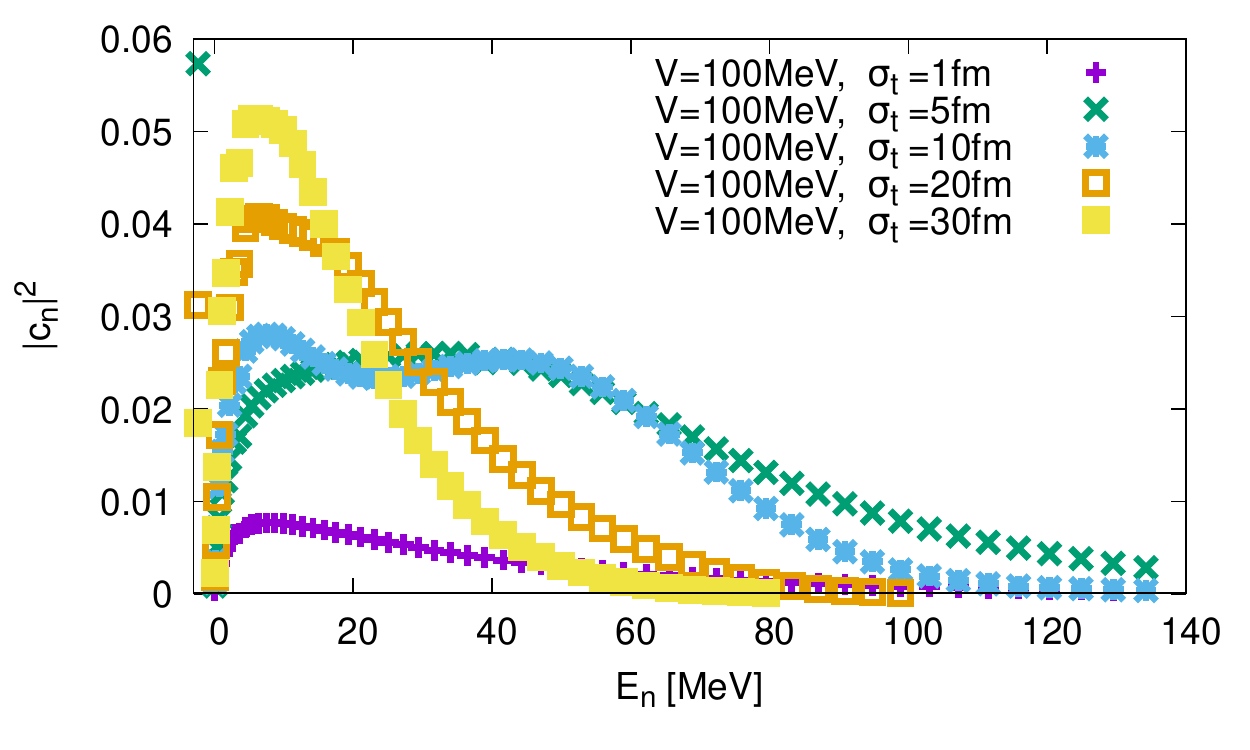}
  	\caption{Distribution of $|c_n(t_\text{final})|^2$, comparing
          different $\sigma_t$ and $V$. $\sigma_x=1.2$ fm, with initial
          condition $|c_0(t=0)|^2 =1$.}
  	\label{fig:distribution}
  \end{figure}
 \begin{figure}
	\centering
	\includegraphics[width=\columnwidth]{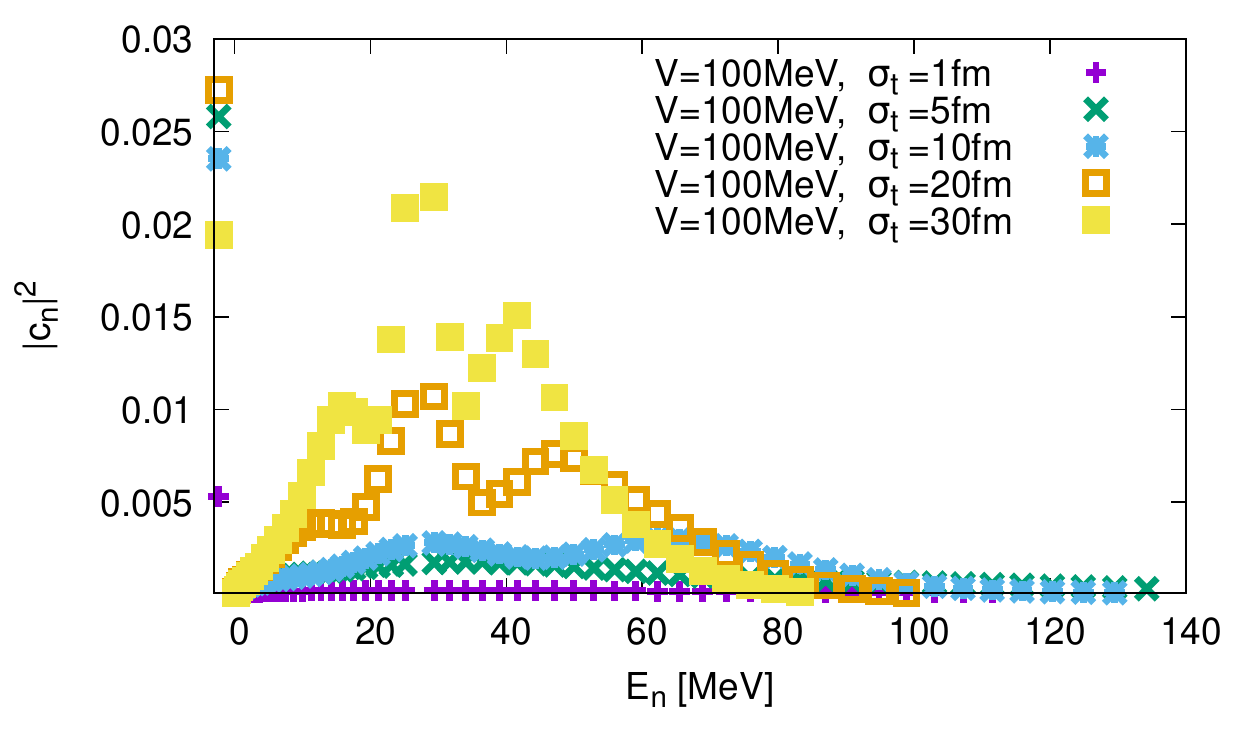}
	\caption{Distribution of $|c_n(t_\text{final})|^2$, comparing different $\sigma_t$ and $V$. $\sigma_x=1.2$ fm, with initial condition $|c_{50}(t=0)|^2 =1$.}
	\label{fig:distribution_c50}
\end{figure}
Heisenberg's energy uncertainty relation,
 \begin{equation}\label{Heisenberg}
 \Delta E \Delta t \geq \frac{\hbar}{2},
 \end{equation}
 should be interpreted as
 \begin{equation}
\label{Heisenberg2}
\Delta E \cdot \sigma_t \geq \frac{\hbar}{2},
 \end{equation}
 where $\Delta E$ is the standard deviation of the energy of the final
 distribution of states, as illustrated in \cref{fig:distribution} and
 \cref{fig:distribution_c50}, and $\sigma_t$ is the standard deviation
 of the pulse length. Let us therefore consider the yellow dots, which
 represent a pulse of length $\sigma_t=30\,\fm$. The expectation value
 for the energy lies here at $\approx 9$ MeV and the width of the
 distribution is $\approx 12$ MeV. Therefore we obtain an energy spread
 of $\Delta E\approx 24$ MeV. If $\sigma_t=30\,\fm$, we obtain
 \begin{equation}
   \Delta E \sigma_t \geq \frac{\hbar}{2} \leftrightarrow 24 \text{ MeV}
   \cdot \frac{30\text{ fm}}{197\text{ MeV fm}} \approx 3.65 >
   \frac{1}{2}, 
 \end{equation}
 which shows, that the interaction time could still be enlarged, to
 obtain a narrower distribution. For smaller $\sigma_t$, for example
 $\sigma_t=1, 5, 10$ fm, more states are excited and the energy
 distribution gets broader. This can also be seen in the strength of the
 excitation, which means the amplitude in $\vert c_n(t\gg
 t_0)\vert^2$. The longer the interaction time of the potential with the
 system, the more a certain state is populated and the more the
 distribution peaks at a certain point.  If one increases $\sigma_t$,
 then the distribution shifts to the left. We want to mention here, that
 those states, that were originally prepared in \cref{fig:distribution}
 and \cref{fig:distribution_c50} are above the range of the axis of
 abscissas in both figures. This is due to the fact, that for small
 $\sigma_t$ the destruction of the initial state is not significant. The
 width of each peak in the distribution is never less than the limiting
 value of $\frac{\hbar}{\sigma_t}$, in accordance with Heisenberg's
 uncertainty relation in energy and time. Therefore we can conclude,
 that the Heisenberg's uncertainty relation in energy and time
 describes, which energy states are occupied after a perturbation, but
 not how fast.  As illustrated in \cref{fig:distribution}, for
 $\sigma_t=10$ fm, and especially in \cref{fig:distribution_c50} there
 are two or three peaks at certain energies in the distribution, which
 correspond to further excitations of the systems of higher orders,
 similar to the harmonics of a string, corresponding to the energy
 differences $\hbar \omega_{mn}$.

\section{Conclusions and Outlook}
\label{sec:Conclusions}
To address the question of bound-state formation and dissociation under
the influence of time-dependent perturbations we have investigated a
simple one-dimensional model of a particle in a square-well potential,
which has been adjusted such that only one bound state exists (analogous
to the case of the deuteron). After solving the energy eigenproblem we
have solved numerically the time evolution of the system under the
influence of Gaussian time-dependent pulses.
We have demonstrated, that disturbing the system with such a pulse leads
to the excitement of higher energy states and a de-excitement of the
bound state, if the bound state is originally fully populated. On the
other hand a before unoccupied bound state can be formed due to the
interaction with the pulse. Due to the large energy
gap, the bound state reacts fastest and evolves in accordance with the
pulse. After the interaction with the potential the state remains
constant. Increasing the number of pulses and applying stochastic pulses
to simulate a thermal bath interacting with the system leads to a
decrease of the originally populated state and an increase of all other
states.

We have investigated the regime in which first-order perturbation theory
is applicable and therefore in a good agreement with the numerical
results. First-order perturbation theory is applicable for short
interaction times, $\sim 1 \,\fm$ and energies $\sim 50 \,\MeV$ to
$\sim150 \, \MeV$. Furthermore, the spacial width of the potential
should be small. We have also shown, that first-order perturbation
theory should be modified in order to obtain a (norm-conserving) master
equation, if one increases the number of pulses.

We have demonstrated, that the system adjusts its states simultaneously
to the potential, independent of the duration of the pulse. We have
shown, that Heisenberg's uncertainty relation for energy and time is
fulfilled in standard deviations of the energy distributions of the
various quantum states and the standard deviation of the duration of the
time-dependent perturbation. Therefore we have seen, that strong and
long pulses interact with the system such, that the potential leads to
certain excitations of energies, which are characterized by the energy
differences, $\hbar \omega_{mn}$.
The width of the distribution is related to Heisenberg's uncertainty
relation in energy and time and does not fall below the limit of
$\Delta E \geq \frac{1}{2\sigma_t}$. On the other hand, short pulses
lead to a weak population of all energy states of the system.
Nevertheless, there is no time delay in the excitation or de-excitation
of a state due to the impact of an interacting potential, but the states
rearrange immediately after the impact of the time-dependent
potential. There is no upper limit of $\braket{E(t)}$, because the time
dependent potential provides the system with additional energy all the
time. 

Since the system is not damped due to a quantum mechanical (dissipative)
Langevin equation the aim of further studies is to develop a
non-equilibrium quantum formalism and evaluate the equilibration of an
open quantum system including distinct bound states, in which our model
serves as a system particle, and dissipation and fluctuation of energy
are introduced by the interaction with a thermal environment. This will
be achieved by employing the influence-functional formalism and the
Caldeira-Leggett master equation in a Lindblad approach \cite{Jan} and
contemporaneously by the use of the Kadanoff-Baym equations \cite{Tim},
to microscopically understand the production of light nuclei, i.e., the
deuteron or heavy quarkonia states.  As a further improvement, the model
will be extended to three dimensions.

\begin{acknowledgments}
  J.R. acknowledges support through the Helmholtz Graduate School for Hadron
and Ion Research for FAIR (HGS-HIRe) and financial support within the framework of the cooperation between
GSI Helmholtz Centre for Heavy Ion Research and Goethe-Universität
Frankfurt am Main (GSI F{\&}E program). We acknowledge support by the Deutsche Forschungsgemeinschaft (DFG, German Re-
search Foundation) through the CRC-TR 211 ‘Strong-interaction matter under extreme conditions’ -- project number
315477589 -- TRR 211. This work is part of a project that has received funding from the
  European Union’s Horizon 2020 research and innovation programme under
  grant agreement STRONG – 2020 - No 824093.
\end{acknowledgments}

\section*{Appendix}
\appendix

\section{Conservation of the norm $\sum_n \vert c_n(t)\vert^2 $}
\label{normconservation}

Let $\psi(t,x)$ be the wave function and
taking the time derivative of the norm,
\begin{equation}
\begin{split}
\label{Schroednorm}
&\frac{\text{d}}{\text{d}t} \int_{-L}^{L} \text{d}x \vert
\psi(x,t) \vert^2 = \int_{-L}^{L}
\text{d}x\frac{\text{d}}{\text{d}t}  \vert \psi(x,t) \vert^2 \\
&= \int_{-L}^{L} \text{d}x \frac{\partial}{\partial t}
(\psi^*(x,t)\psi(x,t)) \\
&= \int_{-L}^{L} \text{d}x\left[ \psi^*(x,t) \frac{\partial
    \psi(x,t)}{\partial t} + \frac{\partial \psi^*(x,t)}{\partial
    t}\psi(x,t) \right].
\end{split}
\end{equation} 
Using the Schr\"odinger equation,
\begin{equation}
\frac{\partial \psi(x,t)}{\partial t} = \frac{\text{i}\hbar}{2m} \frac{\partial^2 \psi(x,t)}{\partial x^2} - \frac{\text{i}}{\hbar} V(x,t) \psi(x,t)
\end{equation}
and its complex conjugate
\begin{equation}
\frac{\partial \psi^*(x,t)}{\partial t} = -\frac{\text{i}\hbar}{2m} \frac{\partial^2 \psi^*(x,t)}{\partial x^2} + \frac{\text{i}}{\hbar} V(x,t) \psi^*(x,t),
\end{equation}
we obtain
\begin{equation*}
\begin{split}
&\frac{\partial}{\partial t}  \vert \psi(x,t) \vert^2 = \psi^*(x,t) \frac{\text{i} \hbar}{2m} \frac{\partial^2 \psi^*(x,t)}{\partial x^2}\\
&-  \frac{\text{i}}{\hbar} V(x,t) \psi(x,t) \psi^*(x,t)\\
&\qquad  - \psi(x,t) \frac{\text{i}\hbar}{2m} \frac{\partial^2 \psi^*(x,t)}{\partial x^2} \\
&+  \frac{\text{i}}{\hbar} V(x,t) \psi(x,t)\psi^*(x,t)\\
&=\frac{\text{i} \hbar}{2m} \left(\psi^*(x,t)\frac{\partial^2 \psi(x,t)}{\partial x^2} - \frac{\partial^2 \psi^*(x,t)}{\partial x^2} \psi(x,t)\right)\\
&=\frac{\partial}{\partial x} \left[\frac{\text{i} \hbar}{2m}
  \left(\psi^*(x,t)\frac{\partial \psi(x,t)}{\partial x} -
    \frac{\partial \psi^*(x,t)}{\partial x} \psi(x,t)\right)\right].
\end{split}
\end{equation*}
This leads, inserting into \cref{Schroednorm} to
\begin{align*}
&\frac{\text{d}}{\text{d}t} \int_{-L}^{L} \text{d}x  \vert \psi(x,t) \vert^2 \\
&= \frac{\text{i} \hbar}{2m} \left. \left(\psi^*(x,t)\frac{\partial \psi(x,t)}{\partial x} - \frac{\partial \psi^*(x,t)}{\partial x} \psi(x,t)\right)\right|_{-L}^{L}.
\end{align*}
With the boundary condition $ \psi(\pm L,t) =0$ it follows
\begin{equation*}
\frac{\text{d}}{\text{d}t} \int_{-L}^{L} \text{d}x  \vert \psi(x,t)
\vert^2 =0.
\end{equation*}
Expanding the wave function,
\begin{align*}
\psi(x,t) = \sum_{n=0}^{m} c_n(t) \psi_n(x),
\end{align*}
also
\begin{align*}
\frac{\text{d}}{\text{d}t} \sum_{n=0}^{m} \vert c_n(t) \vert^2 =0
\end{align*}
holds. This is valid for any $m$, because the ``truncated'' matrix
$H_{jk} = \braket{\psi_j \vert \hat{H}\vert \psi_k}$, with
$j,k \in \left\lbrace 0,1,...,m \right\rbrace$ is Hermitian.

\section{Interpretation of the energy-time uncertainty relation}
\label{Mandelstamm-Tamm}

In this Appendix we briefly discuss the meaning of the ``energy-time
uncertainty relation'', following
\cite{messiah99,Landau1981Quantum,Mandelstam:1945}. Particularly we want
to emphasize that $\Delta t$ does not refer to a kind of ``formation
time'' for bound states in a medium.

It is important to note that the energy-time uncertainty relation needs
a special consideration concerning its interpretation since in quantum
mechanics time cannot be treated as an observable. As has been argued by
Pauli \cite{Pauli:1980_wave_mechanics}, time cannot be interpreted as an
observable in quantum theory since then it would be represented by a
self-adjoint operator, $\hat{t}$, and since the Hamilton operator, which
represents the energy of the system, by definition is the generator of
the time evolution, it had to fulfill the commutation relation
$[\hat{t},\hat{H}]=-\mathrm{i} \hbar$. Then the usual argument familiar
from the analogous situation for position and momentum operators leads
to entire $\R$ as the spectrum for $\hat{H}$. This would imply that the
energy of any system were not bounded from below, i.e., there would be
no ground state of minimal energy and thus matter would not be
stable. In addition it also contradicts the observation of discrete
energy spectra for bound states as in atomic physics.

Now we first consider the usual Heisenberg uncertainty relation for
arbitrary observables $A$ and $B$, represented by the self-adjoint
operators $\hat{A}$ and $\hat{B}$. Let the system be prepared in a pure
state $\ket{\psi}$. For simplicity we define the operators
$\hat{A}'=\hat{A}-\erw{A}$ and $\hat{B}'=\hat{B}-\erw{B}$, where
$\erw{A}=\braket{\psi|\hat{A}|\psi}$ is the expectation value of the
observable $A$. Then the standard deviations $\Delta A$ and $\Delta B$
are given by $\Delta A^2=\erw{A^{\prime 2}}$ and
$\Delta B^2=\erw{B^{\prime 2}}$. Now we define the real quadratic
polynomial,
$f(\lambda)=\braket{(\hat{A}'+\ii \lambda \hat{B}')\psi|(\hat{A}'+\ii
  \lambda \hat{B}')|\psi} \geq 0$. For $\lambda \in \R$ we have
\begin{equation}
\begin{split}
f(\lambda) &= \braket{\psi |(\hat{A}'-\ii \lambda \hat{B}')(\hat{A}'+\ii
  \lambda \hat{B}')|\psi} \\
&= \Delta A^2 + \lambda^2 \Delta B^2 +\erw{\ii
[\hat{A}',\hat{B}']} \lambda \geq 0. 
\end{split}
\end{equation}
Assuming $\Delta B \neq 0$, this is indeed a quadratic polynomial with real
coefficients, and since it is everywhere $f(\lambda) \geq 0$, it can
have at most one real root, which implies that
\begin{equation}
\begin{split}
\label{eq:heisenberg-rob-uncer}
\Delta A^2 \Delta B^2 &\geq \frac{1}{4} \erw{\ii [\hat{A},\hat{B}]}^2 \;
\Rightarrow \\
\Delta A \Delta B & \geq \frac{1}{2} \left |\erw{\ii
    [\hat{A},\hat{B}]} \right|.
\end{split}
\end{equation}
This is the usual Heisenberg uncertainty relation for arbitrary two
observables, $A$ and $B$. It constrains the possibility to
\emph{prepare} quantum states, for which the observables take
well-defined values. Indeed, if the commutator, $[\hat{A},\hat{B}]=0$,
there is no such constraint and there is a complete set common
orthonormal eigenvectors of $\hat{A}$ and $\hat{B}$, i.e., the
observables can simultaneously take precisely defined values. If
$[\hat{A},\hat{B}] \neq 0$, usually such states do not exist, and
preparing the system such that $A$ is rather well defined, i.e.,
$\Delta A$ being small, the value of $B$ is necessarily uncertain, i.e.,
$\Delta B$ must be large. The most famous example is the uncertainty
between the components of the position and momentum of a particle in the
same direction,
\begin{equation}
\Delta x \Delta p_x \geq \frac{\hbar}{2}.
\end{equation} 
Now we consider possible interpretations for an analogous uncertainty
relation between time and energy of a system. Since time, by definition,
is not an observable in quantum mechanics,
\cref{eq:heisenberg-rob-uncer} cannot be directly applied, we have to
specify how to ``measure'' time intervals. This, of course, can only be
achieved by measuring the change of some observable $A$ with time. Since
the Hamiltonian is the generator for the time evolution of the system,
the operator representing the time derivative of the observable $A$ is
\begin{equation}
\label{eq:cov-time-der}
\mathring{\hat{A}}=\frac{1}{\ii \hbar} [\hat{A},\hat{H}].
\end{equation}
Now we can apply the usual uncertainty relation
\cref{eq:heisenberg-rob-uncer} to the energy, i.a., $H$, and the
observable $\hat{A}$ used for time measurement,
\begin{equation}
\Delta H \Delta A \geq \frac{1}{2} \left |\erw{\ii [\hat{A},\hat{H}]} \right| =
\frac{\hbar}{2} \left |\erw{\mathring{\hat{A}}} \right|.
\end{equation}
To resolve a change of $A$ this change should be $\geq \Delta A$, which
means that time intervals measured by observing changes of $A$ with time
have at least an uncertainty
\begin{equation}
\label{eq:energy-time-uncertainty}
\Delta t \geq \frac{\Delta A}{\left |\erw{\mathring{\hat{A}}} \right|}
\geq \frac{\hbar}{2 \Delta H} \; \Rightarrow \; \Delta t \Delta H \geq \frac{\hbar}{2}.
\end{equation}
This means that, the more accurately one likes to measure time intervals
through observation of the change of an observable $A$ with time, the
system used for this measurement must be prepared in a state, for which
the energy uncertainty $\Delta H \geq \hbar/2\Delta t$.

This general consideration can also be applied for the case treated
perturbatively in Sect.\ \ref{heisenberg}. Here the observable, used to
observe the time evolution of the system is the unperturbed energy,
represented by $\hat{A}=\hat{H}_0$, under the influence of the perturbing
external potential $\hat{V}$. Since
$\hat{H}_0=\hat{p}^2/(2m) + V_0(\hat{x})$ and
$\hat{H}=\hat{H}_0+V(\hat{x},t)$ in this case
\begin{equation}
\begin{split}
  \mathring{\hat{A}} = \mathring{\hat{H}}_0 &= \frac{1}{\ii \hbar} [\hat{H}_0,\hat{H}] =
  \frac{1}{2m \ii \hbar} [\hat{p}^2,\hat{V}] \\
&=-\frac{1}{2m} [\hat{p}
  \partial_x \hat{V}+(\partial_x \hat{V}) \hat{p}] = \hat{P},
\end{split}
\end{equation}
where $\hat{P}=(\hat{v} \hat{F}+\hat{F} \hat{v})/2$ with
$\hat{v}=\hat{p}/m$ is the power transferred to the system, described by
$\hat{H}_0$, due to the perturbation $\hat{V}$, i.e.,
$\Delta t=\Delta H_0/\erw{P}$. The corresponding energy-time uncertainty
relation (\ref{eq:energy-time-uncertainty}) holds at any time and for
any state the system is prepared in. 

It is also consistent with the perturbative derivation of the
energy-time uncertainty relation in Sec.\ \ref{heisenberg}, i.e., the
width of the energy distribution for a perturbation of a finite duration
$\Delta t \simeq \sigma_t$ since after the perturbation (or for $t
\rightarrow \infty$ for our Gaussian time dependence of the
perturbation, which has to be understood as an ``adiabatic-switching
procedure'') $\hat{H}=\hat{H}_0$, i.e., then $\Delta H =\Delta
H_0$. Here $\Delta t \simeq \sigma_t$, because only during the time the
perturbation is effective, $\erw{H_0}$ can change. In other words, the power
\begin{equation}
\erw{\hat{P}} \simeq \frac{\Delta H_0}{\sigma_t}
\end{equation}
in accordance with (\ref{Heisenberg2}).

\bibliography{paper.bib}

\end{document}